\documentclass{article}

\usepackage{aas_macros}

\usepackage[english]{babel}
\usepackage[section]{placeins}

\usepackage[letterpaper,top=2cm,bottom=2cm,left=3cm,right=3cm,marginparwidth=1.75cm]{geometry}

\usepackage{amsmath}
\usepackage{graphicx}
\usepackage[colorlinks=true, allcolors=blue]{hyperref}
\usepackage{authblk}

\usepackage[T1]{fontenc}

\usepackage{natbib}

\usepackage{longtable}

\date{June 30, 2022}

\begin{document}
\title{An inflationary disk phase to explain extended protoplanetary dust disks}

\author[1]{Raphael Marschall}
\author[1]{Alessandro Morbidelli}

\affil[1]{CNRS, Observatoire de la Côte d'Azur, Laboratoire J.-L. Lagrange, CS 34229, 06304 Nice Cedex 4, France}

\maketitle

\begin{abstract}
   {Context: Understanding planetesimal formation is an essential first step to understanding planet formation. 
   The distribution of these first solid bodies will drive the locations where planetary embryos can grow, eventually leading to fully-fledged planets.}\\
   \indent{Aim: We seek to understand the parameter space of possible protoplanetary disk formation and evolution models of our Solar System.
   A good protoplanetary disk scenario for the Solar System must meet at least the following three criteria: 1) It must produce an extended gas and dust disk (e.g., ~45 au for the dust); 2) within the disk, the local dust-to-gas ratio in at least two distinct locations must sufficiently increase to explain the early formation of the parent bodies of non-carbonaceous and carbonaceous iron meteorite; and 3) dust particles, which have condensed at high temperatures (i.e., calcium–aluminium-rich inclusion, CAIs), must be transported to the outer disk.
   Though able to satisfy a combination of these three criteria, current protoplanetary disk models have not been successful in recreating all three features.
   We aim to find scenarios that satisfy all three criteria.
   }\\
   \indent{Methods: In this study, we use a 1D disk model that tracks the evolution of the gas and dust disks.
   Planetesimals are formed within the disk at locations where the streaming instability can be triggered.
   We explore a large parameter space to study the effect of the disk viscosity, the timescale of infall of material into the disk, the distance within which material is deposited into the disk, and the fragmentation threshold of dust particles.
   }\\
   \indent{Results: We find that scenarios with a large initial disk viscosity ($\alpha>0.05$), relatively short infall timescale ($T_{\texttt{infall}}<100-200$~kyr), and a small centrifugal radius ($R_C\sim0.4$~au; the distance within which material falls into the disk) result in disks that satisfy all three criteria for a good protoplanetary disk of the Solar System.
   The large initial viscosity and short infall timescale result in a rapid initial expansion of the disk, which we dub the \textit{inflationary phase} of the disk.
   Furthermore, a temperature-dependent fragmentation threshold, which mimics that cold icy particles break more easily, results in larger and more massive disks.
   This in turn, results in more ``icy'' than ``rocky'' planetesimals. 
   Such scenarios are also better in line with our Solar System, which has small terrestrial planets and massive giant planet cores.
   Finally, we find that scenarios with large $R_C$ cannot transport CAIs to the outer disk and do not produce planetesimals at two locations within the disk.}
  {}

\end{abstract}

\section{Introduction}\label{sec:introduction}
Understanding planetesimal formation within protoplanetary disks is an important first step to understanding planet formation.
The distribution of these first solid bodies will drive the locations where planetary embryos can grow, eventually leading to fully fledged planets \citep[e.g.,][]{Chambers2001Icar,Walsh2011Natur}.

Observations of protoplanetary dust disks show two distinct properties: they are large and long-lasting.
Their sizes range from $10-500$~au with typical sizes $\sim30$~au \citep{Tripathi2017ApJ,Andrews2018ApJ,Hendler2020ApJ}, and have lifetimes of millions of years \citep[e.g., ][]{Barenfeld2017ApJ,RuizRodriguez2018MNRAS}.
Because the disk formation occurs on much shorter timescales (of the order of 100 thousand years), dust is not continuously supplied to the system. 
It, therefore, needs to be preserved at large heliocentric distances for millions of years after disk formation.

The Solar System provides a set of additional constraints on the properties and evolution of the protosolar disk. 
However, it is unknown a priori whether these were common to most protoplanetary disks or specific to our own.

The existence and the properties of comets suggest that the protosolar disk was typical in terms of radial extension and lifetime. In fact, 
comets are thought to have formed at distances between 20 and 40~au \citep{Nesvorny2017ApJ,Nesvorny2018ARA&A}. Furthermore, cold classical Kuiper belt objects are thought to have formed in-situ up to a distance of $45$~au \citep{Nesvorny2022AJ}. Additionally, comets have likely formed late \citep[e.g.,][]{Nakashima2015E&PSL,Nimmo2018SSRv,Neumann2018JGRE}, i.e., several million years after the formation of the first solids, the so-called calcium–aluminium-rich inclusion \citep[CAIs;][]{Amelin2010E&PSL,Connelly2012Sci}. A late formation is needed to avoid any significant radiogenic heating, which would result in the loss of highly volatile ices such as $\mathrm{CO_2}$ and CO \citep[e.g.,][]{Eberhardt1987A&A,Morse2015A&A,Gasc2017MNRAS}. The presence of these highly volatile species also in very large comets ($\sim 100$~km) such as Hale-Bopp or Bernardinelli–Bernstein \citep{Capria2000A&A,Kelley2022ApJL} confirms that comets remained cold not because of their small sizes but rather because of they formed late, at a time when most short-lived radioactive elements (e.g. $^{26}$Al) had already decayed. Also, radioactive heating would have increased the bulk density of large objects to a degree inconsistent with the low density of icy bodies such as Trojans and Kuiper-belt objects (between $300$ and $1500$~km/m$^{-3}$; \citealt{Preusker2017,Groussin2019,Berthier2020,Spencer2020}), further supporting late formation.

We have additional evidence for a long-lasting protosolar disk.
The meteoritic record contains both samples from differentiated and un-differentiated parent bodies.
The latter formed significantly later -- up to 5 million years after CAI formation \citep[][]{Nimmo2018SSRv}.
Therefore, ample evidence suggests that our Solar System formed from an extended and long-lived protoplanetary disk.
Because we will focus in this work on the first generation of planetesimal, and the problem of long-lasting disks is an issue in itself, our first requirement for a good model of the Solar System disk is its large size.

Focusing on the first generation of planetesimals, the differentiated parent bodies of iron meteorites, we find that these can be divided into two isotopically distinct groups akin to carbonaceous chondrites (CC) and non-carbonaceous chondrites (NC) \citep{Warren2011, Kruijer2017}.
Thus, they are usually referred to as CC- and NC-iron meteorites, respectively. 
Both groups of iron meteorites formed essentially simultaneously in the disk \citep{Spitzer2021E&PSL}.
Because they have formed simultaneously, they must form at distinctly different locations in the disk that can have a different disk composition.
Therefore, our second requirement for a good model of the Solar System disk is that it produces planetesimals at two distinct locations in the disk.

Finally, the oldest Solar System solids, CAIs, are thought to have formed as high temperature condensates very close (few tenths of an au) to the proto-Sun \citep{Scott2003TrGeo}.
The age of CAIs sets what is usually considered time zero of Solar System formation \citep[see review by][]{Chaussidon2015GMS}.
Their age is $4,567.30 \pm 0.16$ million years according to Pb-Pb dating \citep{Jacobsen2008E&PSL, Connelly2012Sci, Bouvier2010NatGe}.
Recent work argues for a revised age for CAIs of 4,568.7~Myr \citep{Piralla2023Icar, Desch2022arXiv}.
The duration of CAI formation appears to be very short, from $\sim 100$~kyr \citep{Connelly2012Sci} to just $\sim 10$~kyr \citep[][]{Jacobsen2008E&PSL}.
Importantly, the abundance of CAIs is significantly higher in CCs than NCs \citep{Scott2003TrGeo}, the latter of which are thought to have formed closer to the Sun than the former \citep{Warren2011}.
Furthermore, CAIs have even been found in comets \citep{Brownlee2006Sci, Zolensky2006Sci}, which descend from planetesimals formed the farthest away from the Sun.
Therefore, even though CAIs were formed close to the Sun, the planetesimals formed the furthest away are more enriched with them.
This implies that these high-temperature condensates have been transported efficiently to the outer disk, so that the latter became enriched with CAIs while the inner disk remained depleted in CAIs.
The fact that the isotopic compositions of differentiated/early and undifferentiated/late planetesimals overlap within the CC and NC reservoirs, respectively \citep{Kruijer2017} indicates that this division of a CAI-rich outer and CAI-depleted inner disk was present already at the time when the parent bodies of the iron meteorites formed.
It has been proposed that CAIs were transported ballistically to the outer disk via magnetised winds \citep{Shu2001ApJ}.
But modern simulations reveal that only particles much smaller than observed CAIs can be efficiently transported this way \citep{Rodenkirch2022A&A}.
Thus, the radial transport of CAIs during the outward spreading of the disk \citep{Jacquet2011A&A, Pignatale2018ApJL} remains the best option.

In summary, for our Solar System, a disk formation and evolution scenario must satisfy at least the following three properties:
\begin{enumerate}
    \item it must develop an extended disk of gas and dust (up to 45~au for the dust);
    \item in at least two distinct locations in the disk, the dust/gas ratio must be able to increase sufficiently to produce planetesimals and explain the early formation of NC- and CC-iron meteorite parent bodies;
    \item particles which condensed at high temperatures (i.e., CAIs) must be able to reach large heliocentric distances, i.e., be transported from the star's proximity to large distances.
\end{enumerate}

In this work, we try to build such a scenario. In section~\ref{processes}, we describe the key processes in the formation of the disk, the evolution of its gas and dust components and planetesimal formation. Then  we describe the disk model we use (Sec.~\ref{sec:model}) before discussing the model setup (Sec.~\ref{sec:parameters}).
In particular, we will describe four assumptions' influence on satisfying the Solar System constraints.
These are i) the centrifugal radius, $R_C$; ii) the initial viscosity of the disk, $\alpha_0$; iii) the infall timescale of material onto the disk, $T_{\text{infall}}$; and iv) the effect of a temperature-dependent fragmentation threshold for icy particles.
Our results are presented in Sec.~\ref{sec:results}.

We will show that an initial rapid expansion -- forming an inflationary disk stage -- can result in large dust disks, forming planetesimals at two locations in the disk and transporting CAIs to the outer disk.
We will also show that disks forming from clouds with large angular momentum, which readily solves the problem of dust-disk sizes by delivering material directly at large distances, are unable to form planetesimals at two distinct locations and don't allow the transport of CAIs into the outer disk.

\section{Key processes in disk evolution and planetesimal formation}
\label{processes}

As anticipated in the introduction, we start discussing key processes in the formation and evolution of the disk and planetesimal accretion, focusing on the unknowns we will parametrise and test in our models. 

\subsection*{Accretion of material into a protoplanetary disk}
Whether protoplanetary disks are ``born'' big (i.e., form from the outside in) or ``grow up'' to be big (i.e., grow from the inside out) depends on the angular momentum of the infalling material.
Thus, the angular momentum of the pre-stellar cloud determines where material falls into the disk.
The larger the angular momentum of the material, the larger the distance at which it falls into the disk.
The radius in the disk where the angular momentum of the infalling material is equal to the angular momentum of the Keplerian disk is called the centrifugal radius, $R_C$.
If, e.g., the pre-stellar cloud has a constant angular speed throughout, then shells of material closer to the centre collapse first and, having a small specific angular momentum, will fall very close to the proto-star.
More distant shells fall into the disk later and, having larger specific angular momentums, fall farther away from the star.
Therefore, $R_C$ increases with time for a pre-stellar cloud with a constant angular frequency.
Depending on the pre-stellar cloud, the centrifugal radius can be as large as 100~au \citep[e.g.,][]{Shu1977ApJ, Hueso2005A&A, Pignatale2018ApJL}.

However, it is also possible the material falls continuously close to the star because of magnetic braking, which removes a significant amount of the angular momentum of the infalling material \citep[][magnetically braked material flows along the disk surface towards the proto-star, sketched in Fig.~19]{Lee2021A&A}.
The formation of such small disks is observed in some magnetohydrodynamics (MHD) simulations of the gravitational collapse of pre-stellar clouds \citep[e.g.,][]{Machida2019ApJ, Vaytet2018A&A, Machida2011MNRAS}. These disks can then spread radially due to viscous evolution.

Current cloud collapse simulations do not yet provide a firm prescription on how a disk forms and where it collects the material falling from the molecular cloud. 
Thus, in the following, we will test different idealised recipes to identify which best fits the constraints enumerated in the introduction.

Observations suggest that the timescale of accretion of material into the disk is of the order of $10^5$~y, with large uncertainties \citep{Larson1969MNRAS, Vaytet2018A&A, Wurster2021MNRAS, Wurster2022MNRAS}, so that the infall timescale can be considered a free parameter within an order of magnitude. 
Late accretion through streamers is sometimes observed \citep{Tobin2010ApJL, Yen2019ApJ, Pineda2020NatAs} but, given the stochastic nature of this process, we don't include it in our investigations.

The viscosity plays a key role in the evolution of the disk and its spreading away from $R_C$. 
There is a big discussion in the literature on the actual viscosity of protoplanetary disks, but it concerns isolated accretion disks. 
As long as the disk is accreting material from the molecular cloud, it is expected to suffer strong Raynold stresses that act as an effective viscosity \citep{Kuznetsova2022ApJ}. 
Thus, it seems legitimate to assume that a disk which is still accreting mass has a viscosity proportional to the mass infall rate, but the proportionality factor is poorly constrained, and therefore we will consider different values in our study.

\subsection*{Motion of dust particles within the disk}
For disks forming with a small $R_C$ where, e.g., the material never falls outside of 10~au, dust particles must be efficiently transported from the vicinity to distances far away from the star in order to build the large observed dust disks. 
In such cases, the disk (dust and gas) forms from the inside out.
The outward motion of the dust is induced through the radial aerodynamic drag of the radially expanding gas \citep[e.g.,][]{Yang2012M&PS}.
Gas within $R_C$ has a negative radial velocity (towards the star), but the gas close to and beyond $R_C$ viscously spreads outwards.
Eventually, the entire gas disk becomes an accretion disk with a negative radial velocity throughout the disk.

The radial motion of the dust depends on its size.
The important parameter for dust dynamics is not the particle size but its Stokes number, defined as:
\begin{equation}\label{eq:St}
    {\text St} = \frac{\pi a \rho_d}{4 \Sigma_g}    \quad,
\end{equation}
where $a$ is the diameter of the dust particle, $\rho_d$ is the particle solid density, and $\Sigma_g$ is the gas surface density.
The radial dust velocity, $v_r^d$, can then be written as
\begin{equation}\label{eq:radialDustSpeed}
    v_r^d = \frac{2{\text St}}{1+{\text St}^2}v_t^g + \frac{1}{1+{\text St}^2} v_r^g  \quad,
\end{equation}
where $v_t^g$ and $v_r^g$ are the tangential and radial velocities of the gas relative to a circular Keplerian orbit, respectively.
When there is no dust feedback onto the gas, $v_t^g=\eta v_K$ is the difference between the azimuthal gas speed and the Keplerian speed due to the partial pressure support of the gas. 
The radial velocity of the gas is due to viscosity.

For small dust, when ${\text St} \ll 1$, the radial dust speed is dominated by the radial gas speed ($v_r^d \propto v_r^g$, Eq.~\ref{eq:radialDustSpeed}).
Thus, when the dust is small, it initially expands outwards from $R_C$ with the gas.
Once the dust has grown sufficiently (i.e., ${\text St} \sim 1$), the tangential speed of the gas can become the dominant factor in Eq.~\ref{eq:radialDustSpeed}.
Because the gas is sub-keplerian $v_t^g < 0$, the radial dust speed can also become negative once the dust has grown large enough, even if the gas is still in radial expansion.
This reflects the fact that dust particles that are large enough feel the headwind of the gas -- the dust is moving at keplerian while the gas is at sub-keplerian speed.
Thus, while the gas can further expand outwards viscously, large dust particles will begin to drift back towards the star.

\subsection*{Dust growth}
Particles grow on a timescale $1/Z\Omega$, where $Z=\Sigma_d/\Sigma_g$ is the local column integrated dust-to-gas ratio, but their growth is limited by the so-called fragmentation barrier \citep{Drazkowska2017A&A}. When particles grow, they start to partially decouple from the gas. 
The turbulence in the disk and the radial drift of particles in the disk then enhance the relative speeds among dust particles and when the latter is larger than the fragmentation velocity $v_{\text{frag}}$, dust particles cannot coagulate further but rather break upon collisions.

The largest Stokes number that particles can acquire by coagulation is estimated to be \citep{Drazkowska2017A&A} the minimum between:
\begin{equation}
    {\text St}_{\text{frag}} = \frac{0.37 v_{\text{frag}}^2}{3 \texttt{Sc}~ \alpha c_s^2}    \quad,\label{fraglimit}
\end{equation}
and
\begin{equation}
  {\text St}_{\text{ddf}} = \frac{0.37 v_{\text{frag}}}{2 |\eta v_K|}  \quad,
  \label{ddflimit}
\end{equation}
where $\alpha$ is the gas viscosity parameter, following the assumption that the viscosity $\nu=\alpha c_s^2/\Omega$ \citep{Shakura1973A&A}, $ \texttt{Sc}$ is the Schmidt number relating viscous angular momentum transfer to turbulent diffusion, and $c_s$ is the local sound speed. Eq.~(\ref{fraglimit}) comes from the velocity dispersion due to turbulence in the disk, and Eq.~(\ref{driftlimit}) comes from the differential radial speed of particles of different Stokes numbers. 

The fragmentation velocity $v_{\text{frag}}$ depends on the material properties. 
Following the results of laboratory experiments \citep[e.g.,][]{Dominik1997ApJ,Wada2007ApJ,Blum2008ARA&A,Teiser2009MNRAS,Guettler2010A&A}, it is typically assumed that $v_{\text{frag}}=100$~cm/s for refractory and silicate particles whereas $v_{\text{frag}}=1,000$~cm/s for icy particles beyond the water snowline. 
Yet, recent laboratory experiments have shown that ice particles are only 'sticky' close to the sublimation temperature and more brittle when the ice is cold \cite[e.g.,][]{Musiolik2019ApJ}. 
Therefore, we will explore an additional fragmentation threshold prescription for icy particles, which is temperature dependent. 
Similarly, it may be possible that silicate particles become more sticky when their temperature is close to sublimation \citep{Pillich2021A&A} but, awaiting experimental confirmations, we don't yet consider this possibility in our model.

\subsection*{Planetesimal formation}
The currently favoured mechanism for planetesimal formation is through the streaming instability \citep[SI;][]{Youdin2005ApJ, Johansen2007Natur, Johansen2014prpl, WahlbergJansson2014A&A, WahlbergJansson2017MNRAS, Simon2017ApJL, Yang2017A&A, Abod2019ApJ} and subsequent gravitational collapse to form large -- the preferred size of 100~km -- planetesimals \citep[e.g.,][]{Simon2016ApJ, Schafer2017A&A, Klahr2020ApJ, Polak2023ApJ}. 
The SI is triggered once sufficient dust collects within a certain region of the disk and causes the local dust-to-gas ratio to reach some threshold value \citep[e.g., 0.5;][]{Gole2020ApJ}.
At that point, clouds of dust particles collapse under their own gravity to form planetesimals \citep[e.g.,][]{Klahr2020ApJ, Nesvorny2021PSJ, Polak2023ApJ}.

Previous models exploring the formation of planetesimals within a disk have focused on static disks, i.e., snapshots of a given disk phase.
Such models have been successful in showing that planetesimal formation is particularly favoured in the vicinity of sublimation lines, in particular, the water snowline \citep[e.g.,][]{Saito2011ApJ, Ida2016A&A, Drazkowska2017A&A, Hyodo2019A&A, Hyodo2021A&A}.
More recently, these static models were extended to include the temporal evolution of the gas and dust disks and confirm that planetesimal formation at the snowline remains the dominant location for forming a first generation of planetesimals \citep{Drazkowska2017A&A, Drazkowska2018A&A, Charnoz2019A&A, Morbidelli2022NatAs}.
Such evolving disk models capture the expansion phase of the disk and therefore do not rely on a prescribed disk profile, e.g., the surface density of gas and dust.
The addition of the silicate condensation line, in conjunction with a small centrifugal radius, was shown by \cite{Morbidelli2022NatAs} to result in planetesimals forming at the silicate line in addition to those forming at the snow line. 

Yet, these newer, explicitly time-dependent inside-out formation models exhibit the problem that they cannot satisfy at least two of our requirements.
These disks typically don't result in extended disks (requirement 1), and by extension, will also struggle to bring CAIs to the outer disk (requirement 3). This shows that a more in-depth investigation is needed, which motivates the present paper.

The reason why the published models fail on requirements 1 and 3 is that the resulting dust disk sizes are merely slightly larger than the location of the water snowline ($\sim5$~au).
This is because particles beyond the snowline rapidly grow and drift back towards the proto-star on much shorter time scales due to aerodynamic drag in the tangential direction \citep[e.g.,][]{Takeuchi2002ApJ, Takeuchi2005ApJ}.

Thus, the underlying problem is one of the particle sizes and their associated dynamical timescales.
Indeed, equation~\ref{eq:radialDustSpeed} tells us that when the dust growth timescale is much shorter than the timescale for particles to be dragged outwards by the gas, dust will be lost into the star efficiently.
Therefore, to prevent dust particles from drifting towards the star, we must prevent them from growing to large sizes too fast.

\section{Model}\label{sec:model}
We use the previously presented \texttt{DiskBuild} protoplanetary disk model of \cite{Morbidelli2022NatAs}, which includes dust and gas evolution.
Here we summarise the model's main features and refer the reader to the methods section of  \cite{Morbidelli2022NatAs} for a detailed model description.
We only detail the improvements made for this work.

\begin{figure}
	\includegraphics[width=\textwidth]{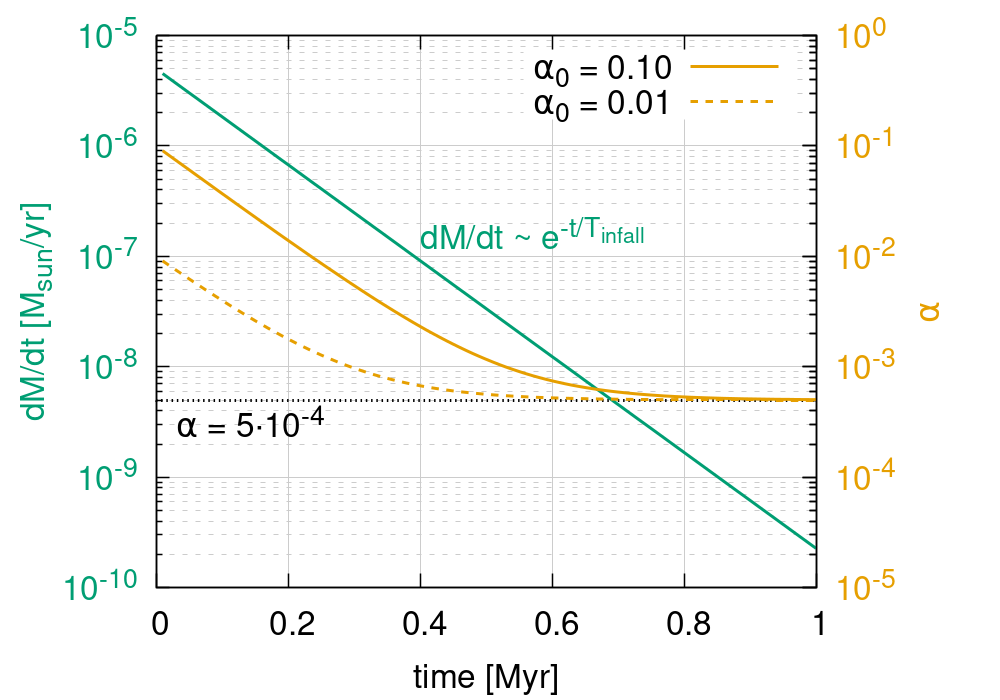}
	\caption{The mass added to the disk as a function of time is shown in green and units of solar masses per year. In this example, the infall timescale, $T_{\text{infall}}=100$~kyr. The yellow lines show the temporal evolution of the viscosity, $\alpha$, for the two end-member cases where $\alpha_0=10^{-1}$ and $\alpha_0=10^{-2}$.}
	\label{fig:mdot-alpha} 
\end{figure}

We typically initiate the model with an empty disk and a proto-star with an initial mass of $0.5M_{\odot}$.
This is consistent with a Class-0 protostar.
Subsequently, the disk is populated through an infall function describing the amount of mass added to the star-disk system as a function of time and distance to the star.
The mass added to the disk is assumed to decay over time as $\exp(-t/T_{\text{infall}})$, where $t$ is time and $T_{\text{infall}}$ is the infall timescale, a free parameter of the model.
The time-integrated mass of the infall is scaled to result in a star-disk system with one solar mass.
The green line in Figure~\ref{fig:mdot-alpha} shows an example of the disk mass infall function for $T_{\text{infall}}=100$~kyr.

\begin{figure}
	\includegraphics[width=\textwidth]{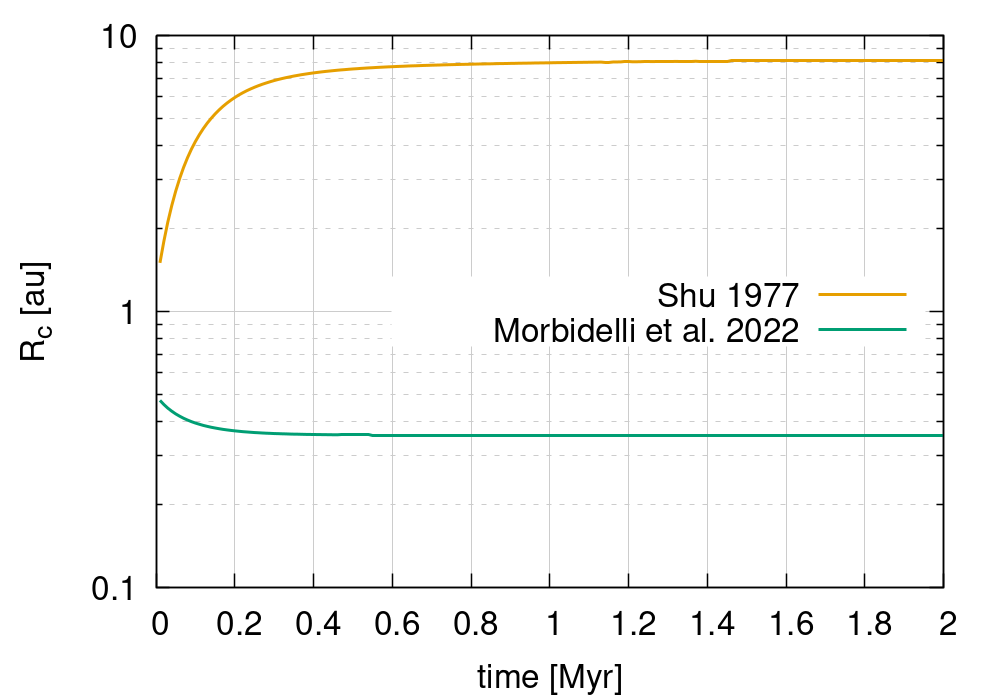}
	\caption{The centrifugal radius, $R_C$, is shown as a function of time for the two cases in this study. The orange line shows the prescription according to Eq.~\ref{eq:RcShu} \citep[$\omega=9\times10^{-15}$~s$^{-1}$ and $T=15$~K;][]{Shu1977ApJ}, assuming that the angular momentum of infalling material increases rapidly with time. This prescription, in addition to one where $R_C$ grows to 100~au, has been used for results in Sec.~\ref{sec:Shu}. The green line shows the function of Eq.~\ref{eq:RcMorbidelli} \citep{Morbidelli2022NatAs} describing an infall scenario where the infalling material loses angular momentum due to magnetic braking. The latter was used for most cases presented in this work.}
	\label{fig:Rc} 
\end{figure}

The maximum distance within which material falls into the disk is the centrifugal radius, $R_C$.
As recalled in section~\ref{processes}, the classic recipe for the evolution of $R_C$ over time is derived from the assumption of a rigidly rotating sphere of material \citet{Shu1977ApJ} and is \cite{Hueso2005A&A}:
\begin{equation}\label{eq:RcShu}
    R_C(t) \simeq 53 \left( \frac{\omega}{10^{-14}\text{s}^{-1}}\right)^2 \left( \frac{T}{10\text{K}} \right)^{-4} \left( \frac{M(t)}{1 M_{\odot}} \right)^3 \text{au}\quad,
\end{equation}
where $\omega$ is the angular speed of the cloud, $T$ is the cloud temperature, and $M(t)$ is the total mass of the star-disk system.
For $\omega=9\times10^{-15}$~s$^{-1}$ and $T=15$~K, $R_C$ and never exceeds $10$~au (orange line in Fig.~\ref{fig:Rc}).
For a larger angular speed of, e.g., $\omega=3.1\times10^{-14}$~s$^{-1}$ the centrifugal radius will grow to $100$~au.
Therefore, depending on the angular speed of the molecular cloud the centrifugal radius can become very large.
As a reference \cite{Pignatale2018ApJL} used $\omega=1\times10^{-14}$~s$^{-1}$ and $T=15$~K for their study bringing $R_C$ to $10.5$~au.

\cite{Morbidelli2022NatAs} suggested that the alternative scenario, where $R_C$ remains small throughout the infall process due to magnetic braking of the infalling material, should be appropriate, at least for our Solar System, to aid the formation of planetesimals at two locations within the disk.
We thus adopt the prescription of \cite{Morbidelli2022NatAs} of
\begin{equation}\label{eq:RcMorbidelli}
    R_C(t) = \frac{0.35}{\sqrt{M_\star(t)}} \, \text{au} \quad,
\end{equation}
where $M_\star$ is the mass of the proto-star in solar masses, $M_\odot$.
We stress that the crucial assumption of Eq.~\ref{eq:RcMorbidelli} is not its exact form but that $R_C$ remains small, particularly that it remains smaller than the condensation line of silicates and refractories.

There is an ongoing debate over the scale at which this disk forms \citep[e.g.,][]{Machida2014MNRAS, Masson2016A&A} and we thus don't constrain ourselves to only exploring scenarios using Eq.~\ref{eq:RcMorbidelli}. Thus, although we mainly present results using that prescription from \cite{Morbidelli2022NatAs} we will also examine the effects of using the more traditional ``Shu recipe'' (see results in Sec.~\ref{sec:Shu}). In particular, we will show results where the $R_C$ grows to 10 and 100~au, respectively.
The prescription of $R_C$ forms our first main assumption in the model.
Material falling closer than $0.05$~au (the inner edge of our simulation domain) is assumed to be directly accreted onto the star.

The gas disk evolves under viscous heating and spreading.
We use the usual definition of the viscosity $\nu=\alpha H^2 \Omega$ (or, equivalently $\nu=\alpha c_s^2 /\Omega)$, where $\Omega$ is the keplerian frequency and $H=\sqrt{\frac{RTr^3}{\mu G M_\star M_\odot}}$ is the scale height, with $R$ the gas constant, $\mu$ the mean molecular weight of the gas, and $G$ the gravitational constant.
The scale height is computed self-consistently at each distance, $r$, of the disk by measuring the temperature, $T$.
The viscosity parameter, $\alpha$, is a free parameter and varies in time and with radial distance.
As discussed in Sect.~\ref{processes}, it is reasonable to assume that $\alpha$ decays over time in a manner proportional to the disk infall function (two examples are shown in Fig.~\ref{fig:mdot-alpha}). 
However, the initial value of $\alpha$ -- denoted $\alpha_0$ -- is considered a free parameter.
A minimum value of $\alpha$ is set at $5\times10^{-5}$, the order of magnitude of the effective turbulence generated by hydrodynamical mechanisms such as the vertical shear instability \citep{Kumar1993MNRAS, Urpin1998MNRAS}.
In addition, at locations in the disk where it is gravitationally unstable or close to instability, the
disk develops clumps and waves that also generate an effective viscosity.
We take this into account by increasing $\alpha$ in those locations locally \citep[see Eq. 8ff in ][]{Morbidelli2022NatAs}.

Of the infalling mass, $1\%$ is considered dust and the rest gas (hydrogen), corresponding to the solar metallicity \citep{Asplund2009ARA&A}.
The dust is further split up into three sub-species: 1) all refractory species with a sublimation temperature above $1,400$~K, 2) silicates with a sublimation temperature of $1,000$~K, 3) water/ice with a sublimation temperature of $170$~K.
In reality, the sublimation temperature for silicates depends on the disk pressure and global chemistry (e.g. the C/O ratio). 
For instance, \cite{Morbidelli2020E&PSL} showed that the silicate sublimation temperature could be 1,060~K for $P=10^{-4}$ bar and C/O=1.0.
For simplicity, we have kept the sublimation temperature of silicates at $1,000$~K.
The species are assumed to have a relative abundance of 0.35/0.35/0.3.
When the local disk temperature is above one of these sublimation temperatures, the corresponding dust specie is considered to be in the gaseous form and thus evolves in the same way the overall gas does.

In the part of the disk where a dust specie is in solid form, we track the size of dust particles, or rather its stokes number, ${\text St}$.
The model has only one dust size at each radial distance, as in most codes.
For dust size distributions that are dominated by the largest size, this is a good approximation and is indeed the result of dust growth models \citep[e.g.,][]{Birnstiel2012A&A, Paruta2016A&C, Mattsson2020MNRAS, Stammler2022ApJ}.

Because of the Eulerian nature of our code, we don't just consider the limit Stokes number given by the fragmentation barriers (\ref{fraglimit}) and (\ref{ddflimit}), where we assume  $\texttt{Sc}=0.1$ \citep{Morbidelli2022NatAs}, but also need to consider that particles cannot be as large that they immediately drift out of a given cell.
This drift boundary is defined as
\begin{equation}
  {\text St}_{\text{drift}} = 0.055 \frac{\Sigma_d}{\Sigma_g} \frac{r\Omega}{\eta v_K}    \quad,
  \label{driftlimit}
\end{equation}
where $r$ is the radial distance to the star.
The  barriers (\ref{ddflimit}) and (\ref{driftlimit}) are additions to the model compared to the one published in \cite{Morbidelli2022NatAs}, which only considered (\ref{fraglimit}).

The final particle size is determined through the minimum among ${\text St}_{\text{growth}}$, given by the growth algorithm with timescale $Z\Omega$, and ${\text St}_{\text{frag}}$, ${\text St}_{\text{ddf}}$ and ${\text St}_{\text{drift}}$.

We have also improved the dust advection treatment in the code.
For each cell, we now calculate the flux of particles out of the current cell to the lower/upper neighbouring cell based on the respective dust speed at the edge of the cell.
Additionally, we compute the flux of particles from the lower/upper cell to the current cell.
Taking into account all four possible loss/gain contributions is important, in particular, at the water snow line, because there the dust size can significantly change from one cell to the next.
The particles beyond the snow line may drift towards the star, while those within the snow line may still drift away from the star.

The dust surface density is evolved, taking into account advection and diffusion.
The back-reaction from the dust onto the gas is accounted for. 
At each timestep, the midplane volume density of the dust and gas is calculated. 
When the ratio of the two exceeds 0.5, we assume that planetesimal formation can occur via the streaming instability in that ring, removing the dust in excess \citep{Gole2020ApJ}.

\section{Model setups and constraints}\label{sec:parameters}
As described in the introduction, the underlying problem that prevents dust from forming a large disk which extends far beyond the water snowline is that it grows too fast.
We will explore two ways to prevent dust from growing to a size large enough to make it drift towards the star during the expansion phase of the gas disk.

\subsection{Expansion speed of the disk}
First, a more rapid expansion of the gas disk -- which in turn drags the dust particles in the radial direction  when the Stokes number is small (Eq.~\ref{eq:radialDustSpeed}) -- can transport dust into more distant regions of the disk before the dust has a chance to grow significantly.
Faster expansion of the gas disk should manifest when the gas viscosity ($\alpha$) is higher or the infall timescale ($T_{\text{infall}}$) is short.
To explore the effect of these two parameters of our model, we have varied them.

For the viscosity, we have one free parameter, the initial value of $\alpha$ at the beginning of the simulation, denoted $\alpha_0$.
Once $\alpha_0$ is set, it decreases as described in Sec.~\ref{sec:model} proportional to the mass added to the disk.
Because the mass added to the disk decays over time, so will $\alpha$.
We have chosen to vary $\alpha_0$ between $0.01$ and $0.1$ and steps of $0.01$.
The lower limit is consistent with the nominal case presented in \cite{Morbidelli2022NatAs}.
The upper limit might be considered quite high, but \cite{Kuznetsova2022ApJ} showed that for cases where the mass that is added to the disk is a large fraction of the disk mass itself, the disk wide $\alpha$ can reach large values (see their Fig.~8).
In particular, when the infalling mass is on the same order as the disk mass, $\alpha$ reaches values of 0.1.
Such a mass ratio is reached early in our simulations.
Therefore, we believe such a high value of $\alpha_0$ is plausible for a brief period at the beginning of the simulation. Remember that we let our $\alpha$ decay over time at the same rate as the infalling material decays (Fig.~\ref{fig:mdot-alpha}). 

An increased viscosity has the added benefit of increasing the relative velocities between the dust particles and, therefore, their collision speeds.
This results in more fragmentation and, thus, smaller particles, making it easier for the gas to transport the dust to large distances.

Regarding the infall timescale, we have tested nine values of $T_{\text{infall}}$ between $15$~kyr and $630$~kyr.
A logarithmic spacing between cases was used.
In combination with the ten different $\alpha_0$, we arrive at 90 simulations.

\subsection{Fragmentation threshold of the dust}
The second way to ensure particles reach larger distances in the disk is more straightforward.
In our nominal cases, we follow the assumptions of \cite{Morbidelli2022NatAs} and impose a fragmentation threshold of $v_{\text{frag}}=100$~cm/s for refractory and silicate particles and $v_{\text{frag}}=1,000$~cm/s for icy particles beyond the water snowline.

\begin{figure}
	\includegraphics[width=\textwidth]{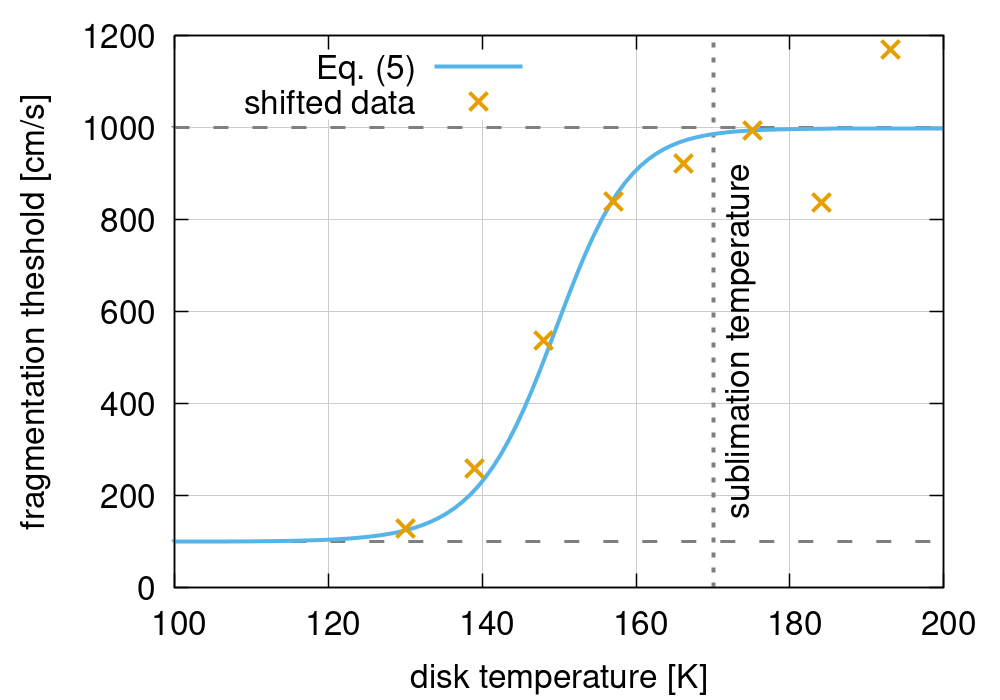}
	\caption{The temperature-dependent fragmentation threshold, $v_{\text{frag}}$, for icy particles according to Eq.~\ref{eq:vLim}} is shown. The experimental data from \cite{Musiolik2019ApJ} have been shifted to lower temperatures to account for the lower sublimation temperature of water in our model compared to the one in the experiment ($\sim220$~K).
	\label{fig:vLim} 
\end{figure}

However, we also test a temperature-dependent fragmentation threshold prescription for icy particles:
\begin{equation}\label{eq:vLim}
    v_{\text{frag}}(T) = v_0 + v_C\Gamma(T)^\frac{5}{6}    \quad,
\end{equation}
where $T$ is the temperature, $v_0=100$~cm/s, $v_C=1,600$~m/s, and
\begin{equation}
\Gamma(T) = \Gamma_C + \Gamma_{d0}\tanh(\beta(T-T_0))    \quad,
\end{equation}
where $\Gamma_C=\Gamma_{d0}=0.25$, $\beta=0.105$, and $T_0=150$.
These parameters ($v_0$, $v_C$, $\Gamma_C$, $\Gamma_{d0}$, and $T_0$) where chosen to match the experimental data presented in \cite{Musiolik2019ApJ}.

Figure~\ref{fig:vLim} shows both the data from \cite{Musiolik2019ApJ} (orange crosses; shifted to account for the different sublimation temperatures between the laboratory and the real disk) and Eq.~\ref{eq:vLim} (light blue line).
The fragmentation threshold decreases from $1,000$~cm/s to $100$~cm/s between disk temperatures of $170$~K and $120$~K.
The new prescription makes icy particles easier to break in cold regions of the disk.
This limits their size and should help transport them to larger distances from the star.

For locations in the disk above the sublimation temperature of $170$~K, i.e., for dry particles, we retain a fragmentation threshold of $100$~cm/s whereas for locations with temperature below $170$~K we use Eq.~\ref{eq:vLim}.
We have run two sets of 90 simulations (the variations in $\alpha_0$ and $T_{\text{infall}}$) for the nominal fragmentation threshold and the new temperature-dependent fragmentation threshold.
The effects from rapidly expanding disks are expected to compound when also applying the new fragmentation threshold.

\subsection{Summary of assumptions}
To summarise, there are four main assumptions that we will explore in this work:
\begin{enumerate}
    \item The centrifugal radius, $R_C$: either according to Eq.~\ref{eq:RcShu} (`Shu recipe') growing to 10 and 100~au respectively or Eq.~\ref{eq:RcMorbidelli} where $R_C$ remains small. Our nominal simulations are performed with Eq.~\ref{eq:RcMorbidelli}.
    \item Variation of the initial disk viscosity, $\alpha_0$, between 0.01 and 0.1.
    \item Variation of the infall timescale $T_{\text{infall}}$ between $15$~kyr and $630$~kyr.
    \item The fragmentation threshold for icy particles: either constant  at 1,000~cm/s (nominal case) or temperature-dependent according to Eq.~\ref{eq:vLim}.
\end{enumerate}

\section{Results}\label{sec:results}

\subsection{Temperature independent fragmentation threshold}\label{sec:resultsTindependent}

\begin{figure*}
	\includegraphics[width=\textwidth]{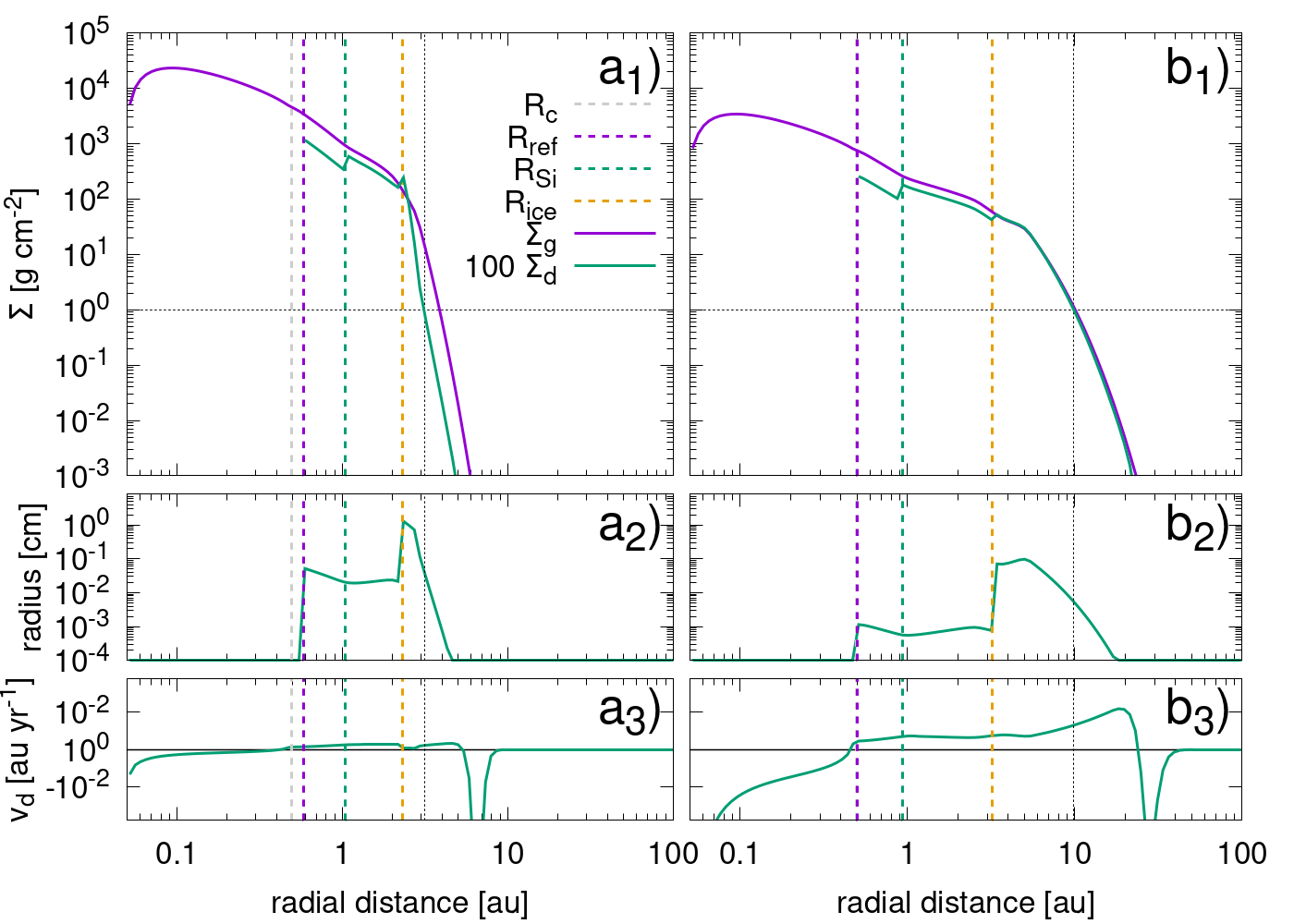}
	\caption{The disk properties for two case with $T_{\text{infall}}=100$~kyr but different initial viscosity (a with $\alpha_0=0.01$ and b with $\alpha_0=0.1$) are shown. Both cases are shown at 1,000~years after disk formation. The top panels (a$_1$ and b$_1$) show the gas surface density (solid purple) and 100 times the dust surface density (solid green) as a function of distance to the star. The vertical dash lines show the centrifugal radius, $R_C$ (grey), the condensation lines of refractories (purple) and silicates (green), and the sublimation line of water (yellow). The fine vertical black line shows the distance where the dust surface density has a value of $10^{-2}$~g/cm$^2$. The middle panels (a$_2$ and b$_2$) show the dust size, and the bottom panels (a$_3$ and b$_3$) the radial dust speed. A positive radial velocity represents motion away from the star, while a negative velocity is towards the star. Both of the shown simulations assume the nominal fragmentation thresholds of 1~m/s and 10~m/s for dry and icy particles, respectively.}
	\label{fig:expansionComparison} 
\end{figure*}

First, we present the results from the cases where the nominal fragmentation threshold for dust particles and the small $R_C$ according to Eq.~\ref{eq:RcMorbidelli} was used.
In these cases, particles within the water snowline fragment at $100$~cm/s while those outside at $1,000$~cm/s.

As discussed in the introduction, the main factor limiting dust transport to large distances is the fast growth and subsequent inward drift of particles once they have crossed the snowline.
Already very early on, e.g., after only 1,000~years, the dust particles just outside the snowline grow to the centimetre scale and effectively stop their outward radial motion.
This is shown in panel a$_1$ of Figure~\ref{fig:expansionComparison}, which depicts the results of the case where we have a small viscosity of $\alpha_0=0.01$ and $T_{\text{infall}}=100$~kyr \citep[nominal case in][]{Morbidelli2022NatAs}.
Particles just outside of the water snowline (dashed yellow line) have a size between 0.1 and 1~cm (Fig.~\ref{fig:expansionComparison}a$_2$) and consequently have almost zero radial velocity (Fig.~\ref{fig:expansionComparison}a$_3$).
Because the gas continues to spread outwards, the dust and gas disks ``decouple'', i.e., the dust expansion lags the one of the gas.
Therefore, even at this very early time, the dust disk is already smaller than the gas disk (fine black dashed line in Fig.~\ref{fig:expansionComparison}).

In contrast, when the initial viscosity is much higher, e.g., $\alpha_0=0.1$ (Fig.~\ref{fig:expansionComparison}b), the dust particles beyond the snowline are roughly an order of magnitude smaller (Fig.~\ref{fig:expansionComparison}b$_2$) and thus retain a positive/outward motion (Fig.~\ref{fig:expansionComparison}b$_3$).
The dust expansion keeps up with the gas expansion, and therefore the two disks retain the same size (Fig.~\ref{fig:expansionComparison}b$_1$). 

As expected, disks with larger viscosity expand faster.
After 1,000 years of expansion, the gas disk with $\alpha_0=0.01$ has expanded to roughly 4~au (measured where the gas surface density is 1~g/cm$^2$).
In contrast, the disk with $\alpha_0=0.1$ has reached 10~au and is, therefore, more than double the size of the other (Fig.~\ref{fig:expansionComparison}b).

\begin{figure*}
	\includegraphics[width=\textwidth]{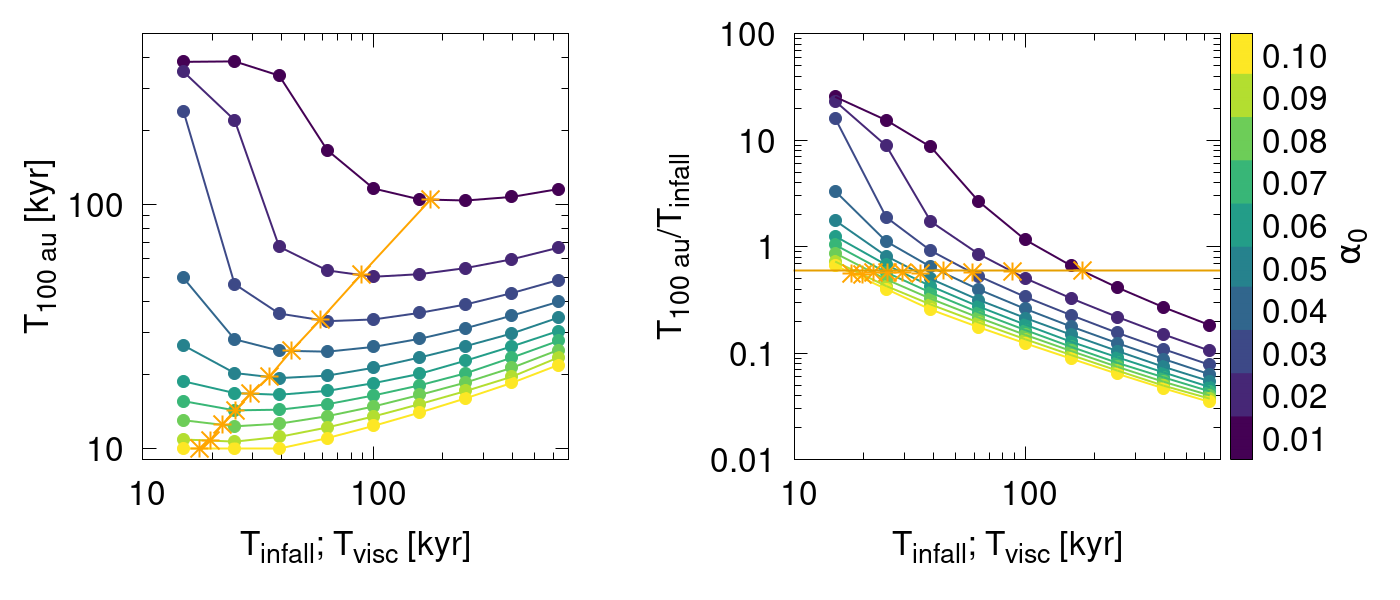}
	\caption{The panes show the time needed for the gas disk to expand to 100~au, $T_{\text{100 au}}$, as a function of the infall timescale, $T_{\text{infall}}$. The size of the disk is measured at the location where the gas surface density is 1~g/cm$^2$. Each curve of different colour represents a different value of the initial viscosity, $\alpha_0$. For each value of $\alpha_0$ the orange star on the corresponding curve indicates the viscous timescale, $T_{\text{visc}}$, of the disk, to be read on the horizontal axis. $T_{\text{visc}}$ represents the average viscous timescale within 10~au at $t=0$ for a disk with an aspect ratio of $6\%$. The left panel shows $T_{\text{100 au}}$ while the right panel shows $T_{\text{100 au}}$ scaled to $T_{\text{infall}}$}
	\label{fig:timescaleForExpansion} 
\end{figure*}

For a given $T_{\text{infall}}$, the time a disk takes to reach 100~au, denoted $T_{\text{100 au}}$, decreases as the initial viscosity increases (Fig.~\ref{fig:timescaleForExpansion}).
To measure the size of the disk, we have used the location where the gas surface density takes a value of 1~g/cm$^2$.
For the dust, we have adopted a value 100 times smaller than the gas because of the metallicity of our infalling material being $1\%$, i.e., a value of 0.01~g/cm$^2$.
We are aware that this choice is somewhat arbitrary but have found it to be the definition that leads to the easiest and most reliable measure of the disk size, particularly for the dust.

Other definitions, e.g., using the distance containing a certain fraction of the total mass, have proven unstable for the dust.

Figure~\ref{fig:timescaleForExpansion} also shows that there is a transition of the expansion regime.
For each value of $\alpha_0$, the orange star on the corresponding curve indicates the viscous timescale, $T_{\text{visc}}$, of the disk, to be read on the horizontal axis. 
$T_{\text{visc}}$ represents the average viscous timescale within 10~au at $t=0$ for a disk with an aspect ratio of $6\%$.
When the infall timescale is shorter than the viscous timescale (on the left side of the orange line), the expansion of the disk slows as the infall timescale decreases.
In the extreme case where the infall timescale is much shorter than the viscous timescale, the disk's ability to spread viscously is limited. 
Thus, the expansion timescale reaches a plateau.
This can be clearly seen in the case of the lowest viscosity case.

In contrast, when the infall timescale is larger than the viscous timescale, the expansion of the disk slows with increasing infall timescale.
This means the expansion is limited by the amount of material resupplied by the infall.
In the most extreme cases $T_{\text{100 au}}\sim400$~kyr (when $\alpha_0$ and $T_{\text{infall}}$ are minimal) and $T_{\text{100 au}}\sim10$~kyr (when $\alpha_0$ is maximal and $T_{\text{infall}}$ is minimal).
We baptise such a rapid expansion, reaching 100~au in just a few tens of thousands of years, the {\it inflationary phase} of the disk.

Because $T_{\text{infall}}$ in these tests varies by more than one order of magnitude, we might better measure $T_{\text{100 au}}$ in units of $T_{\text{infall}}$.
Indeed, the right panel of Fig.~\ref{fig:timescaleForExpansion} shows the expansion time as a fraction of the infall timescale.
In this view, we can recognise that for a given $\alpha_0$, the expansion time as a fraction of $T_{\text{infall}}$ always decreased with increasing $T_{\text{infall}}$.
It is remarkable that if $T_{\text{infall}}=T_{\text{visc}}$, the value $T_{\text{100 au}}/T_{\text{infall}}$ is independent of viscosity (i.e. the orange stars fall on a horizontal line).

\begin{figure*}
	\includegraphics[width=\textwidth]{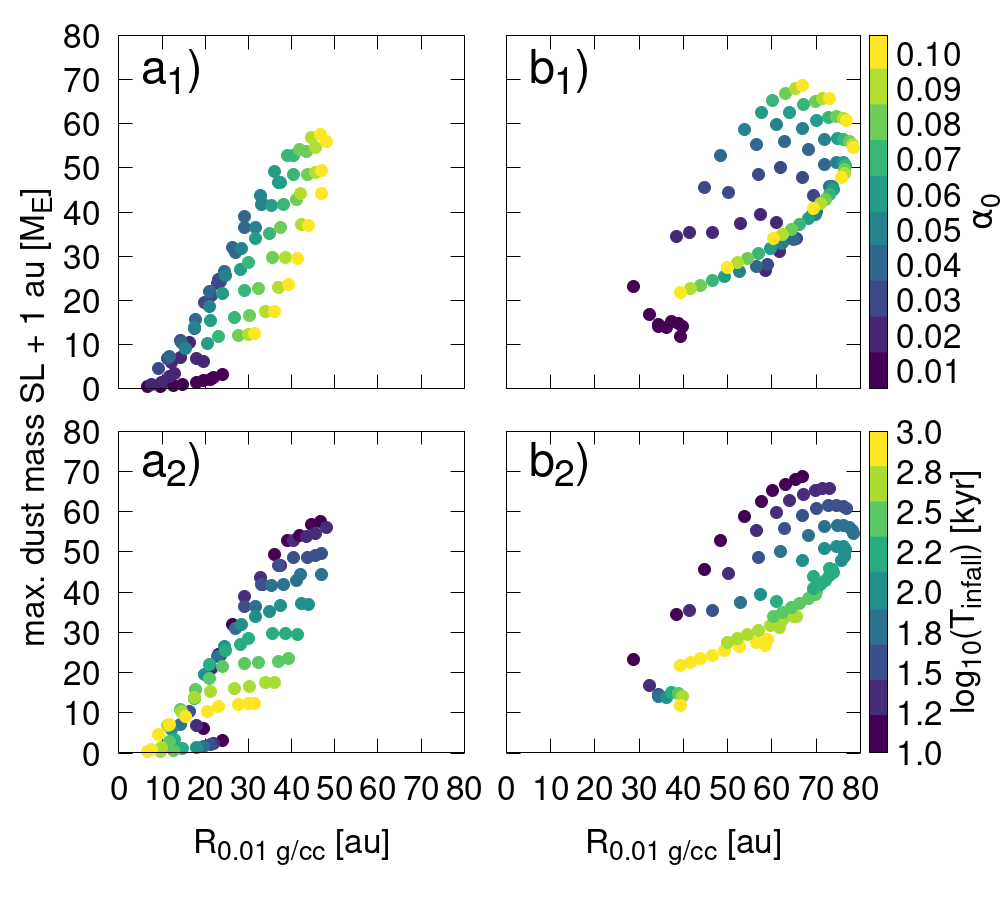}
	\caption{The maximum dust mass 1~au beyond the snowline (SL) is shown as a function of the size of the dust disk, $R_{\text{0.01 g/cc}}$, at the time when the disk has reached that maximum mass. The top panels show the dependency on the viscosity, $\alpha_0$, while the bottom row the dependency on the infall timescale, $T_{\text{infall}}$. Panels a shows the results for the temperature-independent fragmentation threshold, while the b panels show the results for the temperature-dependent fragmentation threshold.}
	\label{fig:maxMass} 
\end{figure*}

\subsubsection{Mass and size of the dust disk}
We have measured the maximum dust mass a given disk holds 1~au beyond the snowline.
To make sure the measurement was not contaminated by the dynamics around the snowline, we chose to exclude the dust mass just outside the snowline.
We will refer to this part of the disk as the `outer disk'.
These masses and sizes are illustrated in Fig.~\ref{fig:maxMass}.

Disks with an initial small viscosity result in small disks that contain little to no dust beyond the snowline (Fig.~\ref{fig:maxMass}a$_1$).
In these cases, the disks can be as small as 5~au.
The most massive disks are formed with the highest viscosity and reach $60$~M$_\oplus$ and sizes between 30 and 50~au.
For a given viscosity, the infall timescale plays a crucial role in determining the dust mass in the outer disk.
The shorter the infall timescale, $T_{\text{infall}}$, is, the more massive the outer disk is (Fig.~\ref{fig:maxMass}a$_2$).
Therefore, short $T_{\text{infall}}$ and large $\alpha_0$ produced the largest and most massive outer disks.
These disks thus satisfy our first criteria for good protoplanetary disks of the Solar System.

\begin{figure*}
	\includegraphics[width=\textwidth]{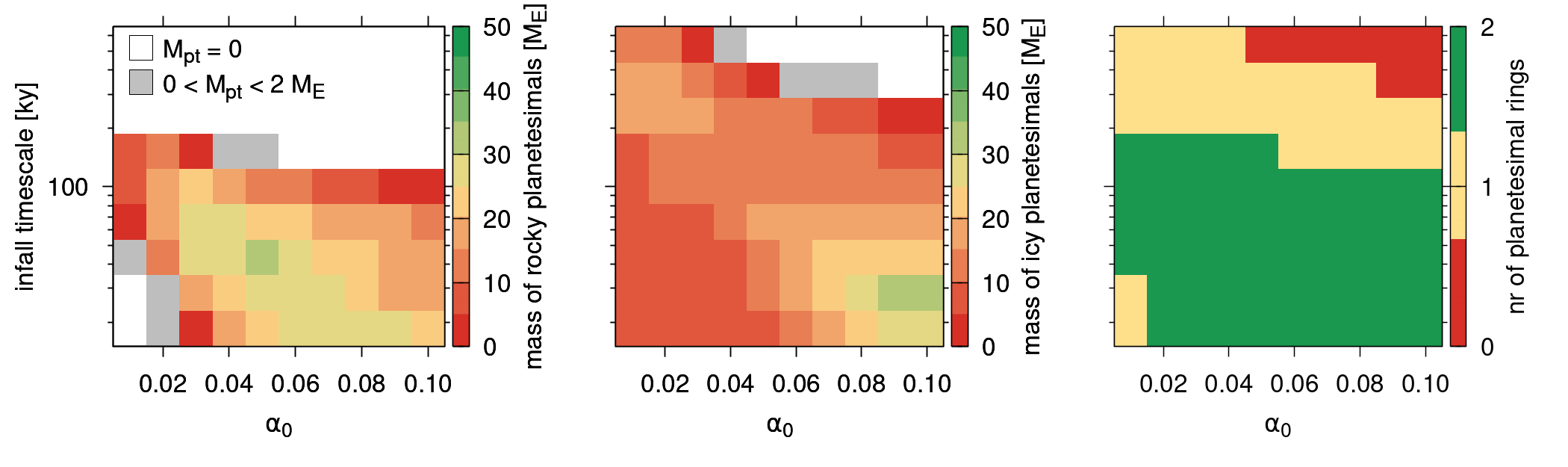}
	\caption{The panels show the mass of planetesimals formed at different locations in the disk. The left panel shows the total mass of ``rocky'' planetesimals as a function of the infall timescale, $T_{\text{infall}}$, and the viscosity, $\alpha_0$. Rocky planetesimals are the ones that form around the silicate condensation line. White areas are disks that don't produce any planetesimals, while grey squares indicate disks that produce between zero and two Earth masses of planetesimals. The centre panel show the total mass of ``icy'' planetesimals for a given disk. Icy planetesimals are the ones formed around the water snowline, typically outside of it. The right panel shows the number of locations, i.e., rings, where planetesimals form.}
	\label{fig:planetesimalsTindependant} 
\end{figure*}

\subsubsection{Planetesimal formation}
To address our second criterion for good protoplanetary disks of the Solar System, we evaluate whether planetesimals form and at how many locations in the disk.
Figure~\ref{fig:planetesimalsTindependant} summarises the mass of planetesimals formed in each of the disks.
Because planetesimals typically form at up to two locations in the disk (Fig.~\ref{fig:planetesimalsTindependant}, right panel), we have split the results into ``rocky'' planetesimals (forming at the silicate condensation line) and ``icy'' planetesimals forming at/outside of the water snowline.

First, we observe that for most cases with $T_{\text{infall}}>100$~kyr, no ``rocky'' planetesimals are formed.
Second, for ``rocky'' planetesimals, there is an optimal viscosity given a $T_{\text{infall}}$.
This is most clearly visible for $T_{\text{infall}}=39$~kys (the third line from the bottom).
For this infall timescale, the optimum viscosity to produce ``rocky'' planetesimals is $\alpha_0=0.05$.
The planetesimal mass decreases for higher and lower values of $\alpha_0$.
When the viscosity is too low, the amount of mass transported to the planetesimal forming region is too small because of the lower radial velocity of the gas, and when the viscosity is too high, the dust cannot settle sufficiently in the midplane to trigger the SI.
Third, the mass of ``icy'' planetesimals is maximised the larger the viscosity and the shorter the infall timescale.
This comes from the fact that those disks are also the most massive beyond the snowline (Fig.\ref{fig:maxMass}a).
Fourth, a small part of our parameter space (high viscosity and long infall timescales) does not form any planetesimals at any location in the disk.
Fifth, the reservoirs of ``rocky'' and ``icy'' planetesimals have a similar order of magnitude in mass.

\subsubsection{CAI transport to the outer disk}\label{sec:CAI}
For the third criterion for good protoplanetary disks of the Solar System, we track high-temperature condensates.
For this purpose, we introduce dust tracers, one for refractory particles that condensate at the refractory line, and a second for refractories that never sublimated.
A fraction of the high-temperature condensates will be CAIs, but in our model, we will just refer to such particles as potential CAIs because we do not track the full condensation sequence of refractories but rather just treat all refractories as one species of dust.
Nevertheless, this lets us determine the locations in the disk that will be enriched or depleted in CAIs.

\begin{figure*}
	\includegraphics[width=\textwidth]{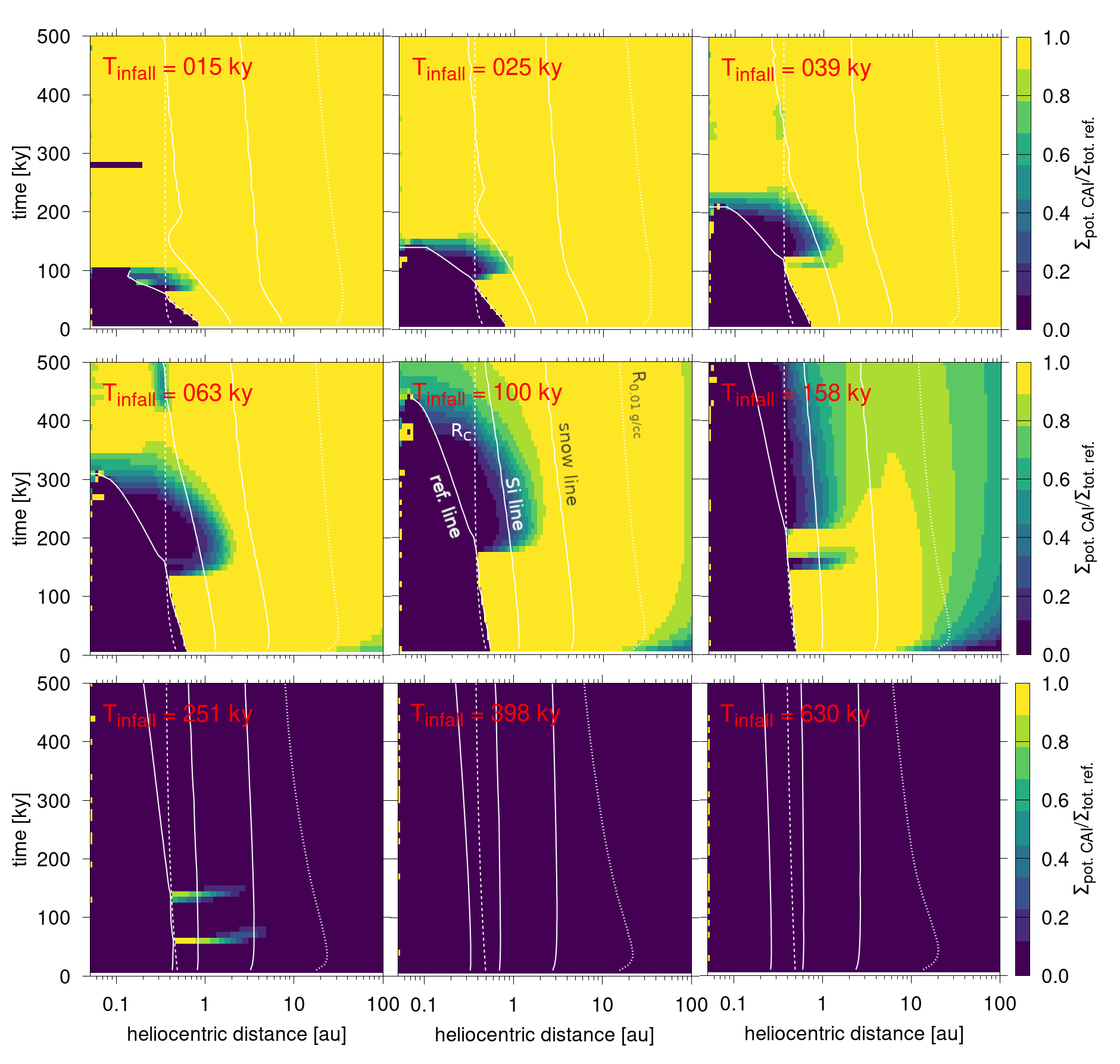}
	\caption{The panels show the ratio between the surface density of high-temperature condensates, $\Sigma_{\text{pot.CAI}}$, and the surface density of all refractory particles, $\Sigma_{\text{tot.ref.}}$ for different values of $T_{\text{infall}}$ and $\alpha_0=0.05$. These simulations assumed the nominal fragmentation thresholds of 1~m/s and 10~m/s for dry and icy particles, respectively. The three solid white lines are, from closest to the star to farthest, the sublimation lines of refractories, silicates, and water. The white course dashed line is $R_C$, and the fine dashed line is the dust disk size, $R_{\text{0.01 g/cc}}$.}
	\label{fig:CAI-Morby-0.05alpha} 
\end{figure*}

The ability of the disk to transport CAIs to the outer disk and retain them there depends again on the viscosity of the disk and the infall time scale.
In particular, the transport of CAIs is promoted when the centrifugal radius is smaller than the refractory condensation line.
If the infall timescale is too long (larger than $\sim 200$~ky for $\alpha_0=0.05$) the disk is rather cold from the beginning, and therefore the refractory condensation line (defined as $T=1,400$~K) is located inside $R_C$, and no CAIs are transported to the outer disk (Fig.~\ref{fig:CAI-Morby-0.05alpha}).
In contrast, when the infall timescale is short (less than $\sim 100$~ky) CAIs are efficiently transported to the outer disk, but then drift back into the inner disk due to the fast evolution of the disk, which transitions to a fully accreting disk within $3-4 T_{\text{infall}}$.
While we show these results for $\alpha_0=0.05$ they are qualitatively the same for other initial viscosities.
For larger initial viscosities, the infall timescale where the disk is too cold to create CAIs is shorter (e.g., at $T_{\text{infall}}\sim 150$~kyr for $\alpha_0=0.1$).
Conversely, this transition happens at larger infall timescales when the viscosity is smaller (e.g., at $T_{\text{infall}} > 400$~kyr for $\alpha_0=0.01$).
But in all cases, neither very short nor long $T_{\text{infall}}$ are favoured for the transport of CAIs to the outer disk.

\begin{figure*}
	\includegraphics[width=\textwidth]{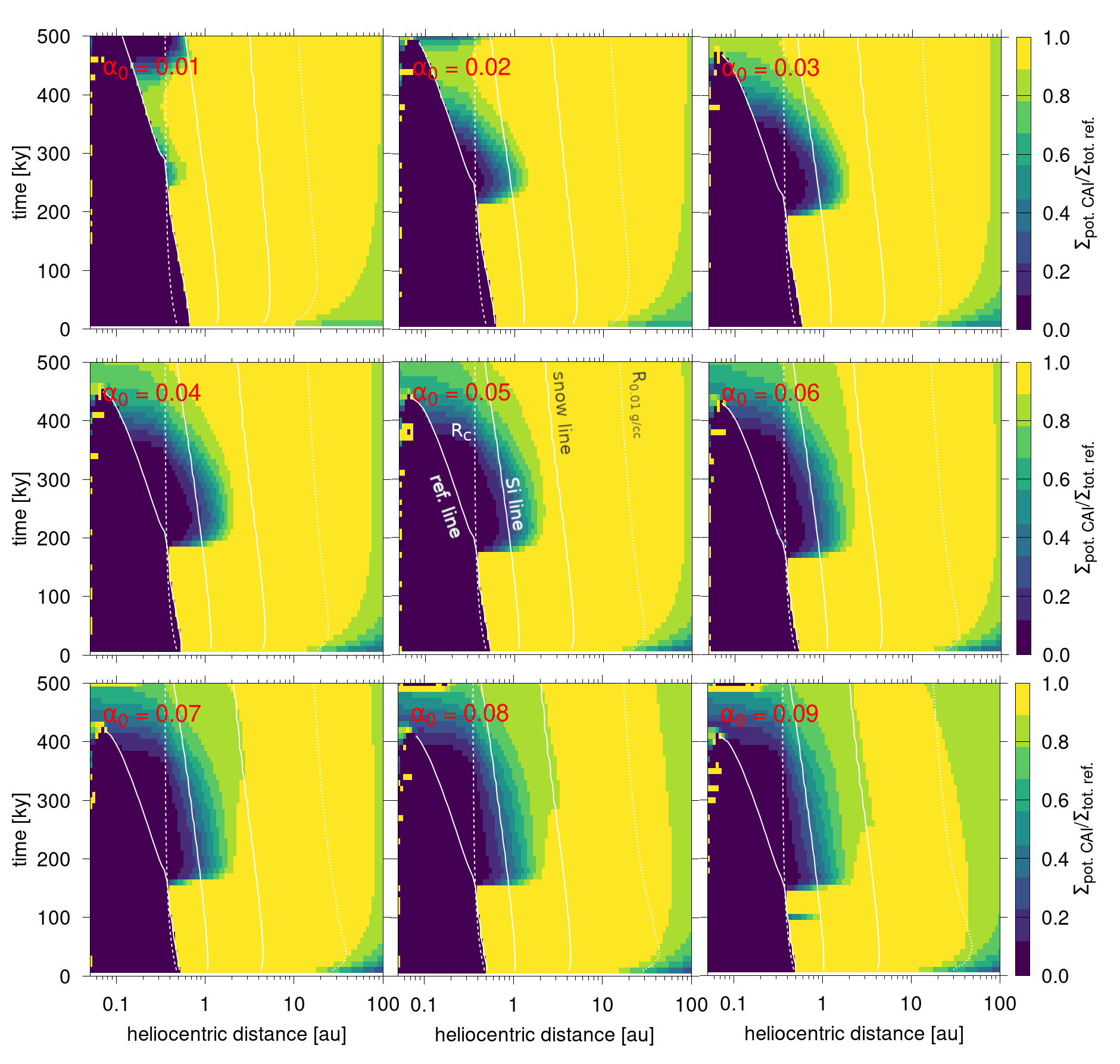}
	\caption{The panels show the ratio between the surface density of high-temperature condensates, $\Sigma_{\text{pot.CAI}}$, and the surface density of all refractory particles, $\Sigma_{\text{tot.ref.}}$ for different values of $\alpha_0$ and $T_{\text{infall}}=100$~kyr. These simulations assumed the nominal fragmentation thresholds of 1~m/s and 10~m/s for dry and icy particles, respectively. The three solid white lines are, from closest to the star to farthest, the sublimation lines of refractories, silicates, and water. The white course dashed line is $R_C$, and the fine dashed line is the dust disk size, $R_{\text{0.01 g/cc}}$.}
	\label{fig:CAI-Morby-100ky} 
\end{figure*}

The smaller the initial viscosity is, the larger the fraction of the disk that is populated by CAIs.
For example, when $\alpha_0 < 0.05$ for $T_{\text{infall}}=100$~kyr the inner disk gets similarly enriched with CAIs as the outer disk (Fig.~\ref{fig:CAI-Morby-100ky}).
When in addition to a low initial viscosity, the infall timescale is also short, then the entire disk is populated by potential CAIs.
Such disks would clearly not match the observations.
Yet, the larger the initial viscosity, the clearer the divide is between a CAI-enriched outer and CAI-depleted inner disk.
The presence of CAIs in outer planetesimals thus suggests a high initial viscosity with the associated rapid expansion phase of the disk.
This appears to be consistent with large, kinetic, Si isotopic variations observed in refractory inclusions, which suggest a turbulent environment during condensation \citep[e.g.,][]{Marrocchi2019PNAS}.

In all of our simulations, we have kept the Schmidt number at $\texttt{Sc}=0.1$.
A higher Schmidt number of, e.g., $\texttt{Sc}=1$ would aid the transport of CAIs to the outer disk.
However, the larger Sc the more the dust will have difficulty settling in the midplane and thus tend to make planetesimal formation more difficult.

\subsection{Temperature dependent fragmentation threshold}\label{sec:resultsTdependent}
In the case where we impose the temperature-dependent fragmentation threshold beyond the snowline (see Sec.~\ref{sec:parameters} and Fig.~\ref{fig:vLim}), we expected that dust fragments more easily and therefore, the outer disk gets populated with more mass.
Indeed, all disks now have at least $10$~M$_{\oplus}$ in the outer disk (Fig.~\ref{fig:maxMass}).
Though the disks are, in general, not significantly more massive (10-70~M$_{\oplus}$ compared to $0-60$~M$_{\oplus}$), the disks with the temperature-dependent fragmentation threshold are much larger ($30-80$~au instead of $5-50$~au).
Thus there is, as expected, a general shift to more massive and larger outer disks.

\begin{figure*}
	\includegraphics[width=\textwidth]{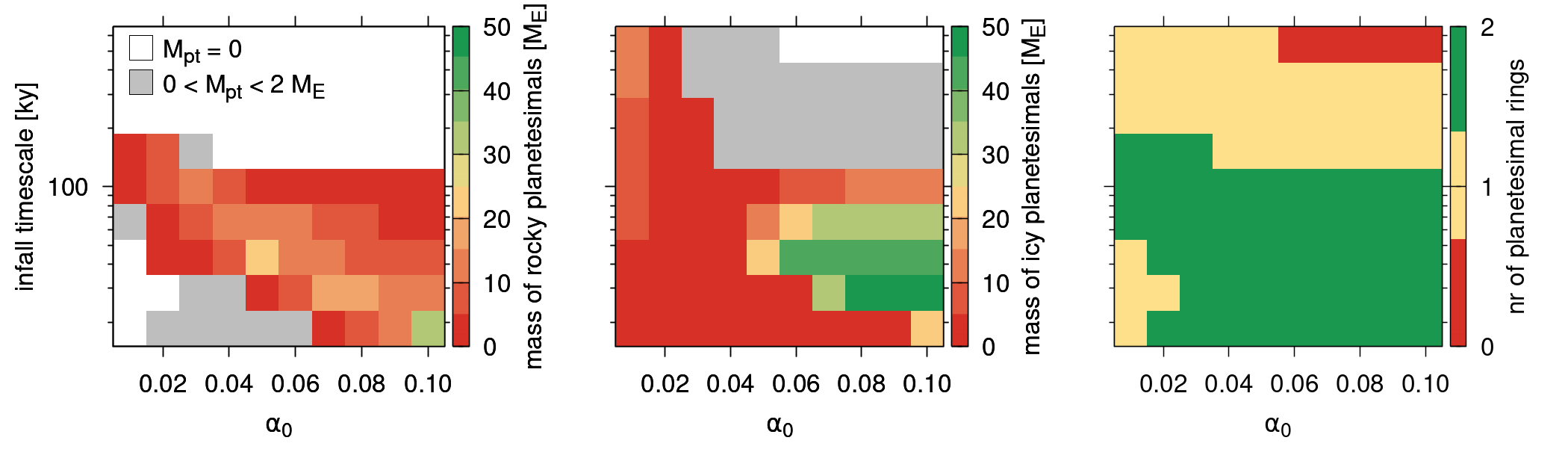}
	\caption{The same panels are shown as in Fig.~\ref{fig:planetesimalsTindependant} but for the temperature-dependent fragmentation threshold.}
	\label{fig:planetesimalsTdependant} 
\end{figure*}

This shift of dust mass from the inner to the outer disk has clear consequences.
We now have significantly more ``icy'' planetesimals than ``rocky'' ones (Fig.~\ref{fig:planetesimalsTdependant}).
For some combination of parameters $\alpha_0$ and $T_{\text{infall}}$ (e.g., $0.07 \le \alpha_0 \le 0.1$ and $40\text{ kyr} \le T_{\text{infall}} \le 100$~kyr), a couple of Earth masses of ``rocky'' planetesimals form together with a couple of tens of Earth masses of ``icy'' planetesimals.
This is in very good agreement with the structure of the Solar System, with massive giant planets' cores and small terrestrial planets.

Similarly to the temperature-independent fragmentation threshold, there are little to no planetesimals when $T_{\text{infall}}>100$~kyr.
The delineation is even a bit clearer.
Nevertheless, the part of parameter space with two planetesimals rings is roughly equally large irrespective of the fragmentation threshold.

Concerning CAI transport, the overall behaviour is similar to the case with the nominal fragmentation threshold.
But, because particles are more easily transported to the outer disk CAIs also reach much larger distances.

\subsection{Shu infall}\label{sec:Shu}

\begin{figure}
	\includegraphics[width=\textwidth]{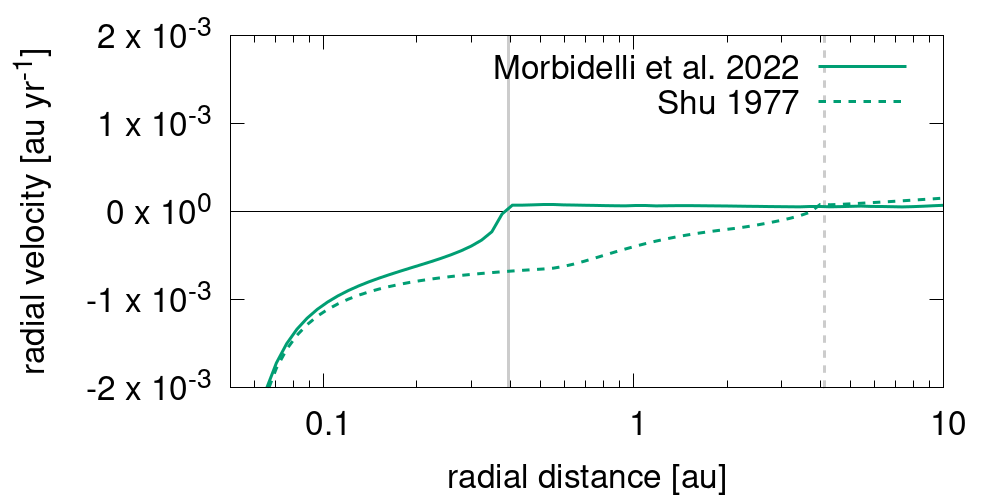}
	\caption{The radial velocity of the gas at $100$~kyr is shown as a function of the radial distance to the star for the case when $T_{\text{infall}}=100$~kry and $\alpha_0=0.01$. The solid line shows the results assuming the infall prescription of \cite{Morbidelli2022NatAs}, while the dashed line is for the ``Shu-infall'' \citep{Shu1977ApJ}. The vertical grey lines denote the position of the centrifugal radius, $R_C$, of the respective case.}
	\label{fig:radialSpeed} 
\end{figure}

Because our prescription of the infall is somewhat unconventional, i.e., the centrifugal radius, $R_C\sim0.35$~au (Eq~\ref{eq:RcMorbidelli}), we have also tested the more common assumption according to \cite{Shu1977ApJ}.
In the ``Shu-case'' the $R_C$ rapidly grows from 1~au to 8~au (Fig.~\ref{fig:Rc}, Eq.~\ref{eq:RcShu} with $\omega=9\times10^{-15}$~s$^{-1}$ and $T=15$~K).
This is because the molecular cloud is assumed to be a rigidly rotating body and angular momentum is conserved (i.e. no magnetic braking).
Therefore, gas with small angular momentum collapses into the disk first, close to the star.
Later, outer shells with larger specific angular momenta fall at larger distances.
This behaviour is in contrast to our preferred cases described above, where magnetic braking reduces the angular momentum of the infalling gas to roughly a fixed value independently of the initial angular momentum of the gas in the molecular cloud.

A major consequence of the ``Shu-type'' infall is connected to the radial gas speed.
The disk within $R_C$ is an accretion disk, i.e., the radial gas velocity, $v_{r,g}$, is negative (Fig.~\ref{fig:radialSpeed}).
Therefore, dust within $R_C$ will also always have a negative radial velocity ($v_{r,d}<0$).
Outside of $R_C$, the disk can spread viscously outwards ($v_r^g>0$; Fig.~\ref{fig:radialSpeed}), and therefore small dust particles will also have a positive radial motion as long as they do not grow large enough to feel the headwind of the gas and start drifting back towards the star.

We have tested two different angular velocities, $\omega$, of the molecular cloud.
Once with $\omega=10^{-14}$~s$^{-1}$ resulting in a maximum $R_C$ of roughly $10$~au as shown in Fig.~\ref{fig:radialSpeed} and once with $\omega=3.1 \times 10^{-14}$~s$^{-1}$ resulting in a maximum $R_C$ of roughly $100$~au.
The temperature of the molecular clouds is assumed to be 15~K in both cases.
We use here the evolution of $R_C$ according to Eq.~3 of \cite{Hueso2005A&A}.
The prescription of the $T_{\text{infall}}$ and $\alpha_0$ remain the same as above.

\begin{figure*}
	\includegraphics[width=\textwidth]{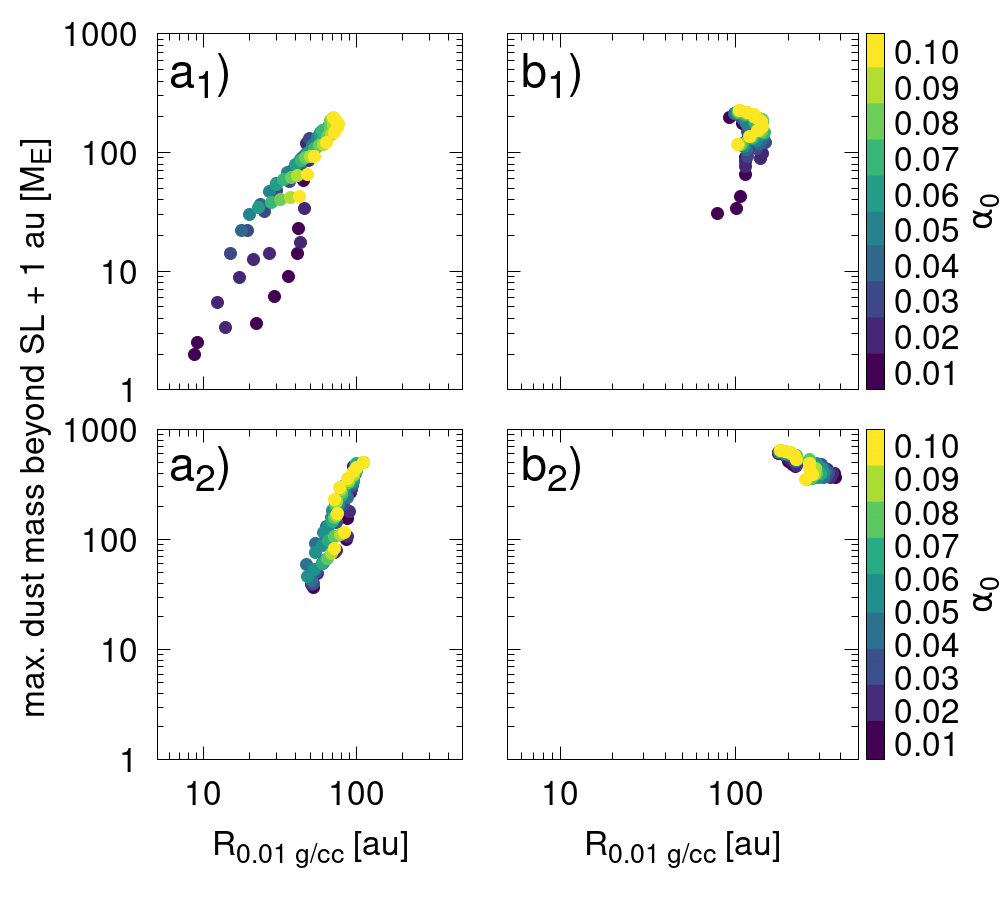}
	\caption{The maximum dust mass 1~au beyond the snowline (SL) is shown as a function of the size of the dust disk, $R_{\text{0.01 g/cc}}$, at the time when the disk has reached that maximum mass. Results for the ``Shu-type'' infall where $R_C$ grows to roughly $10$~au are shown in panels a, while the ones where $R_C$ grows to roughly $100$~au are shown in panels b. Panels with subscript 1 show the results for the temperature-independent fragmentation threshold, while panels with subscript 2 show the results for the temperature-dependent fragmentation threshold.}
	\label{fig:maxMassShu} 
\end{figure*}

In all cases studied the ``Shu-type'' infall has no difficulty producing large and massive disks (Fig.~\ref{fig:maxMassShu}).
When $R_C$ grows to 10~au, and we use the nominal temperature-independent fragmentation threshold, the disks are between 10 and 100~au and have masses between 2 and 200~M$_{\oplus}$ (Fig.~\ref{fig:maxMassShu}a$_1$).
For the same molecular cloud angular velocity but with the temperature-dependent fragmentation threshold, the disks are overall larger and more massive in particular for the cases with small $\alpha_0$. 
The sizes and masses are also confined to 80-150~au and 30-300~M$_{\oplus}$ (Fig.~\ref{fig:maxMassShu}b$_1$).

When $R_C$ grows to 100~au the disks are even larger and more massive.
For the temperature-independent fragmentation threshold, the disks are between 40 and 100~au and have masses between 30 and 600~M$_{\oplus}$ (Fig.~\ref{fig:maxMassShu}a$_2$).
For the temperature-dependent fragmentation threshold, the disk sizes and masses are only weakly dependent on $\alpha_0$ and $T_{\text{infall}}$.
These disks are between 150 and 400~au and have masses between 300 and 700~M$_{\oplus}$ (Fig.~\ref{fig:maxMassShu}b$_2$), and therefore very massive and large.

\begin{figure*}
	\includegraphics[width=\textwidth]{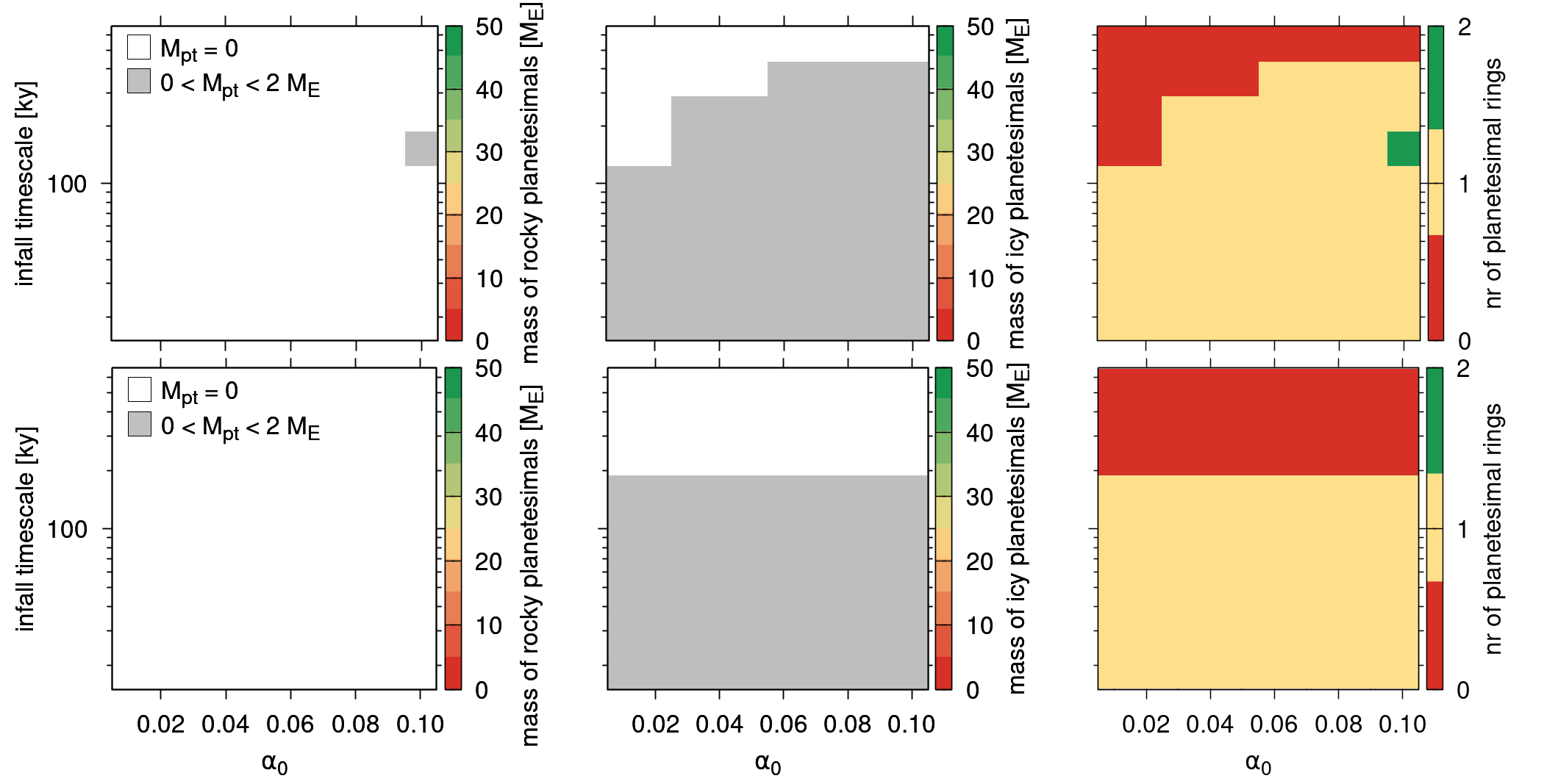}
	\caption{The same panels are shown as in Fig.~\ref{fig:planetesimalsTindependant} but for the temperature-dependent fragmentation threshold and the ``Shu-infall'' model with a maximum $R_C$ of roughly 10~au (top panels) and 100~au (lower panels) respectively. The results for the temperature-independent fragmentation threshold are qualitatively the same. White areas represent cases without any planetesimal production, while grey areas are cases where the planetesimal mass is between zero and two M$_{\oplus}$.}
	\label{fig:planetesimalsShu} 
\end{figure*}

When we prescribe the ``Shu-infall''  particles in the inner disk (within the water snowline) drift rapidly towards the star (Fig.~\ref{fig:radialSpeed}).
This does not allow them to pile up at the silicate sublimation line, and therefore no ``rocky'' planetesimals are formed in any of the cases (left panels in Fig.~\ref{fig:planetesimalsShu}).
Additionally, even at the water snowline, we observe only sparse formation of planetesimals (centre panel in Fig.~\ref{fig:planetesimalsShu}).
This result is largely independent of which angular velocity of the molecular cloud we used and which fragmentation threshold is applied.

Our results differ from the results found by \cite{Drazkowska2018A&A}.
We do not find any planetesimal formation during the phase when the snow line moves outwards.
This might be caused by the different assumptions of the disk infall prescription.
We assume that the mass added to the disk decays over time while a constant function with a sudden cut-off is assumed in \cite{Drazkowska2018A&A}.
Additionally, we find much fewer planetesimals at the snow line.
We believe that \cite{Drazkowska2018A&A} overestimated the amount of water vapour in their disks due to a difference in treatment of the inner disk boundary condition for water vapour to that of hydrogen. 
This supports planetesimal formation.

\begin{figure*}
	\includegraphics[width=\textwidth]{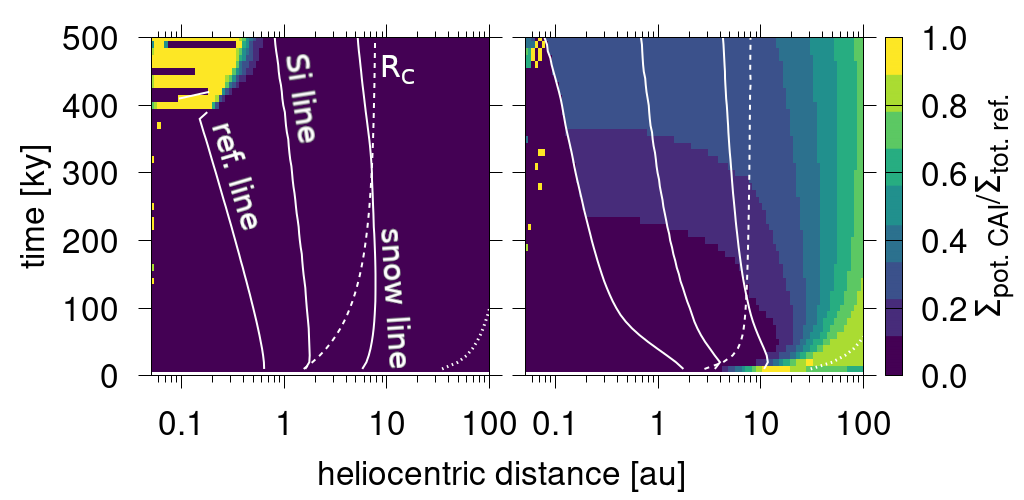}
	\caption{Show the ratio between the surface density of high-temperature condensates, $\Sigma_{\text{pot.CAI}}$, and the surface density of all refractory particles, $\Sigma_{\text{tot.ref.}}$. The case on the left assumes $\alpha_0=0.05$, $T_{\text{infall}}=100$~kyr, and the case on the right assumes $\alpha_0=0.01$, $T_{\text{infall}}=15$~kyr. In both cases, we've used the temperature-dependent fragmentation threshold and the $R_C$ that grows to roughly 10~au. The three solid white lines are, from closest to the star to farthest, the sublimation lines of refractories, silicates, and water. The white dashed line is $R_C$.}
	\label{fig:CAIShu} 
\end{figure*}

Finally, we have also studied the transport of CAIs in such disks.
As expected no CAIs are able to reach the outer disk, or even the terrestrial planet region (Fig.~\ref{fig:CAIShu}).
The example shown in the left panel of Fig.~\ref{fig:CAIShu} assumes $\alpha_0=0.05$, $T_{\text{infall}}=100$~kyr, the temperature-dependent fragmentation threshold, and $R_C$ growing to roughly 10~au but is representative of almost all combinations of $\alpha_0$ and $T_{\text{infall}}$.
The only exception is for $\alpha_0=0.01$ and $T_{\text{infall}}<25$~kyr (right panel of Fig.~\ref{fig:CAIShu}).
In this case, some potential CAIs are produced and transported to the outskirts of the disk (at roughly 100~au).
For cases where $R_C$ grows to roughly 100~au, the situation is even worse because in none of the cases are there any potential CAIs in the disk.
This behaviour is not surprising.
The inward motion of the gas prevents any CAIs from being transported to the terrestrial planet region or outer disk.

Our results are broadly consistent with those of \cite{Pignatale2018ApJL} in that the fraction of CAIs is largest in the outermost part of the disk (towards the edge of the disk itself).
\cite{Pignatale2018ApJL} assume a constant function for the infall of material into the disk, whereas we assume a decaying function.
Assuming a constant source function results in $R_C$ growing much slower than in our cases.
This in turn extends the period during which $R_C$ is smaller than the refractory condensation line.
Therefore, CAIs can be produced for longer and transported into more distant regions of the disk. 
This way the disk generally can be more enhanced with CAIs than in our cases.

\section{Discussion and conclusion}\label{sec:discussion}

Infall of material into protoplanetary disks occurs more or less close to the star -- typically much less than the observed disk sizes).
The disks, therefore, undergo an initial phase of viscous spreading \citep{Lynden-Bell1974MNRAS, Hueso2005A&A}.
The dust on the one hand is entrained in the outward motion of the gas, and on the other hand is slowed down by the sub-keplerian motion of the gas (see Eq.~\ref{eq:radialDustSpeed}) which causes its inward drift.
Whether the radial outward entrainment or sub-keplerian drag dominates the dust motion depends on the particle size.

A key parameter in any protoplanetary disk model is the so-called centrifugal radius, $R_C$.
This is the radius in the disk where the angular momentum is the same as that of the infalling material.
If e.g., the pre-stellar cloud rotates as a rigid sphere \citep{Shu1977ApJ}, then shells of material closer to the centre collapse first and, having a small specific angular momentum, fall very close to the proto-star.
Outer shells, with larger angular momentum, will fall at larger distances and in a later stage in disk formation \citep{Shu1977ApJ}.
In such scenarios, $R_C$ grows with time and we refer to them as ``Shu-type'' infall models.
Contrary to this, magnetic braking can remove angular momentum from the infalling material.
This can cause the material to fall close to the star irrespectively of the initial angular momentum of the material.

In the introduction we have described that a disk formation and evolution scenario for the Solar System must satisfy at least the following three requirements:
\begin{enumerate}
    \item it must develop an extended disk of gas and dust (up 45~au for dust);
    \item in at least two distinct locations in the disk, the dust/gas ratio must be able to enhance sufficiently to produce planetesimals and explain the early formation of NC- and CC-iron meteorite parent bodies;
    \item particles which condensed at high temperatures (i.e., CAIs) must be able to reach large heliocentric distances, i.e., be transported from the star's proximity to large distances.
\end{enumerate}

We found that scenarios using a ``Shu-type'' infall model with an associated large $R_C$ are very successful in achieving requirement 1, as they easily result in large and massive disks.
Yet they fail to produce planetesimals at two locations in the disk (requirement 2) and transport CAIs to the outer disk (requirement 3).
Therefore, these scenarios are bad candidates for the Solar System protoplanetary disk. 
On the other hand, we show that a disk fed by material with a small $R_C$ can satisfy all three requirements, in particular when the initial viscosity is large, the infall timescale is of order or smaller than $100$~kyr.

The main results from our nominal disks with a small centrifugal radius, $R_C$, can be summarised as follows.
\begin{enumerate}
    \item The larger is the initial viscosity, $\alpha_0$, the larger is the outer dust disk.
    \item The shorter is the infall timescale, $T_{\text{infall}}$, the more massive is the outer dust disk.
    \item Therefore, an initial inflationary expansion phase is needed to produce large, massive dust disks. The disk can reach a size of 100~au within a few tens of thousands of years.
    \item A temperature-dependent fragmentation threshold is more realistic and results in significantly larger and slightly more massive dust disks because particles are more fragile and therefore remain smaller at cold temperatures.
    \item No ``rocky'' and very few ``icy'' planetesimals form when $T_{\text{infall}}>100$~kyr.
    \item The largest mass of ``icy'' planetesimals forms when $\alpha_0>0.05$.
    \item There is an optimum $\alpha_0$ that maximises the mass of ``rocky'' planetesimals. For example, for $T_{\text{infall}}=39$~kyr it is $\alpha_0=0.05$.
    \item The temperature-dependent fragmentation threshold results in more ``icy'' than ``rocky'' planetesimals (roughly by a factor of 10) than in the conventional case where the two are of the same order. This is a direct consequence of the temperature-dependent fragmentation threshold resulting in more massive outer disks.
\end{enumerate}

Although our disks with a small $R_C$ can satisfy the three requirements we had put forth at the beginning, there are two additional related requirements that will need to be met eventually but cannot at this point.
Observations show that protoplanetary disks are long-lived, i.e., $3-4$~million years \citep{Andrews2020ARA&A}.
All dust in our models (even in the ``Shu-type'' infall models) drifts into the star on a timescale of a few hundred thousand years.
Therefore, the entire dust disk is lost on that timescale.
Not only does this prevent us from explaining long-lived disks, but our disks are also not able to produce a generation of planetesimals late enough to avoid differentiation, because no dust is available at these later times.
The retention of a large disk and the production of a population of planetesimals that forms late are two additional requirements for a good protoplanetary disk of the Solar System.

Clearly, our model lacks some additional disk processes that can prevent the loss of dust from the disk.
For example, once the disk viscosity is sufficiently small magneto-hydrodynamic (MHD) effects might become dominant and structures (rings and gaps) might appear, impeding dust drift \citep[e.g.,][]{Bethune2016A&A, Riols2020A&A}.
This will be the object of future work.

\section*{Acknowledgments} 
\noindent We acknowledge the funding from the European Research Council (ERC) under the European Union’s Horizon 2020 research and innovation programme (Grant agreement No. 101019380).
Additionally, we acknowledge support from programme ANR-20-CE49-0006 (ANR DISKBUILD).
We thank Sebastien Charnoz, Yves Marrocchi and Francesco Lovascio for reading the manuscript and providing helpful comments.
We thank the anonymous reviewer for their constructive and useful comments that helped us improve the paper.

\bibliography{main}{}
\bibliographystyle{aasjournal}


\end{document}